\documentclass[twocolumn]{aastex631}

\usepackage{import}
\usepackage{xcolor}
\usepackage{comment}
\usepackage{ulem}
\usepackage{booktabs}
\usepackage{xspace}

\graphicspath{ {./Images/} }

\newcommand{\mdot}{$\dot{M}$\xspace}

\newcommand{\mstar}{M$_{\star}$\xspace}
\newcommand{\rstar}{R$_{\star}$\xspace}
\newcommand{\lstar}{L$_{\star}$\xspace}
\newcommand{\dstar}{d$_{\star}$\xspace}
\newcommand{\lacc}{L$_{\mathrm{acc}}$\xspace}
\newcommand{\lphot}{L$_{\mathrm{Phot}}$\xspace}
\newcommand{\lphotexcess}{L$_{\mathrm{Phot,\ Excess}}$\xspace}

\newcommand{\rin}{R$_{in}$\xspace}

\newcommand{\amax}{a$_{max}$\xspace}

\newcommand{\accrateunits}{$\times$10$^{-8}$ M$_{\odot}$ yr$^{-1}$\xspace}

\newcommand{\hst}{\textit{HST}}

\newcommand{\fpeak}{$f_{\mathrm{peak}}$\xspace}
\newcommand{\df}{$\delta f$\xspace}

\defcitealias{RE19}{RE19}

\shorttitle{Accretion Variability in T Tauri Stars. II. Photometric Light Curves}
\shortauthors{Wendeborn et al.}

\begin{document}

\title{A Multi-wavelength, Multi-epoch Monitoring Campaign of Accretion Variability in T Tauri Stars  from the ODYSSEUS Survey. II. Photometric Light Curves}

\correspondingauthor{John Wendeborn}
\email{wendebo2@bu.edu}

\author[0000-0002-6808-4066]{John Wendeborn}
\affil{Institute for Astrophysical Research, Department of Astronomy, Boston University, 725 Commonwealth Avenue, Boston, MA 02215, USA}

\author[0000-0001-9227-5949]{Catherine C. Espaillat}
\affil{Institute for Astrophysical Research, Department of Astronomy, Boston University, 725 Commonwealth Avenue, Boston, MA 02215, USA}

\author[0000-0003-4507-1710]{Thanawuth Thanathibodee}
\affil{Institute for Astrophysical Research, Department of Astronomy, Boston University, 725 Commonwealth Avenue, Boston, MA 02215, USA}

\author[0000-0003-1639-510X]{Connor E. Robinson} 
\affil{Department of Physics \& Astronomy, Amherst College, C025 Science Center 25 East Drive, Amherst, MA 01002, USA}

\author[0000-0001-9301-6252]{Caeley V. Pittman}
\affil{Institute for Astrophysical Research, Department of Astronomy, Boston University, 725 Commonwealth Avenue, Boston, MA 02215, USA}

\author[0000-0002-3950-5386]{Nuria Calvet}
\affil{Department of Astronomy, University of Michigan, 1085 South University Avenue, Ann Arbor, MI 48109, USA}

\author[0000-0001-7157-6275]{\'{A}gnes K\'{o}sp\'{a}l}
\affil{Konkoly Observatory, HUN-REN Research Centre for Astronomy and Earth Sciences, CSFK, MTA Centre of Excellence, Konkoly-Thege Mikl\'os \'ut 15-17}
\affil{Institute of Physics and Astronomy, ELTE E\"otv\"os Lor\'and University, P\'azm\'any P\'eter s\'et\'any 1/A, 1117 Budapest, Hungary}
\affil{Max Planck Institute for Astronomy, K\"onigstuhl 17, 69117 Heidelberg, Germany}

\author[0000-0001-5707-8448]{Konstantin N. Grankin}
\affil{Crimean Astrophysical Observatory, 298409 Nauchny, Republic of Crimea}

\author[0000-0001-7796-1756]{Fredrick M. Walter}
\affiliation{Department of Physics and Astronomy, Stony Brook University, Stony Brook NY 11794-3800, USA}

\author[0000-0003-0292-4832]{Zhen Guo}
\affil{Instituto de F{\'i}sica y Astronom{\'i}a, Universidad de Valpara{\'i}so, ave. Gran Breta{\~n}a, 1111, Casilla 5030, Valpara{\'i}so, Chile}

\author[0000-0001-6496-0252]{Jochen Eisl\"offel}
\affil{Th\"uringer Landessternwarte, Sternwarte5, D-07778 Tautenburg, Germany}

\begin{abstract}

Classical T Tauri Stars (CTTSs) are young, low-mass stars which accrete material from their surrounding protoplanetary disk. To better understand accretion variability, we conducted a multi-epoch, multi-wavelength photometric monitoring campaign of four CTTSs: TW~Hya, RU~Lup, BP~Tau, and GM~Aur, in 2021 and 2022, contemporaneous with $HST$ UV and optical spectra We find that all four targets display significant variability in their light curves, generally on days-long timescales (but in some cases year-to-year) often due to periodicity associated with stellar rotation and to stochastic accretion variability. There is a strong connection between mass accretion and photometric variability in all bands, but the relationship varies per target and epoch. Thus, photometry should be used with caution as a direct measure of accretion in CTTSs.

\end{abstract}


\section{Introduction}
\label{sec: Introduction}

Classical T Tauri Stars (CTTSs) are young ($<$10 Myr), low-mass ($<$2 M$_{\odot}$) stars surrounded by a protoplanetary disk. The disk actively feeds material onto the star along the magnetic fields lines in a process known as magnetospheric accretion \citep[see reviews by][]{Bouvier2007, Hartmann2016}. This accretion process is highly energetic, as fast-moving material shocks near the stellar surface. X-rays produced by the shocks heat the underlying photosphere, creating accretion-induced hotspots that often peak at UV-optical wavelengths \citep{Calvet1998}. The result is significant excess continuum emission at those wavelengths plus strong line emission, both of which can be used to infer properties of accretion \citep[e.g.][]{Ingleby2013, Alcala2017, RE19, Nature}. 

The accretion process is variable and this can be seen in light curves. Previous high-cadence observations of CTTSs \citep{Cody2014, Cody2018, Siwak2018, Robinson2021, Zsidi2022} have revealed that not only are CTTSs variable on timescales from seconds to years, but they also exhibit a wide array of light curve behaviors and morphologies. In many cases, the light curve morphology is indicative of the many sources of variability present. Sources of accretion variability may include (but are not limited to) inhomogenous accretion flows, hotspot plasma oscillations, inner disk inhomogeneity, inner disk thermal instabilities, magnetorotational instability, and gravitational instability \citep[see review by][]{PPVII10}. Other sources of variability include stellar rotation, chromospheric activity, disk occultation, variable extinction, winds/outflows, and flares \citep{Alencar2010, Dupree2013, Cody2014, Hinton2022}.

Due in part to the non-accretion related variability mechanisms, \citet{Robinson2022} show that the connection between accretion and photometric brightness becomes tenuous at longer wavelengths, starting in the optical. The accretion signatures are most closely traced by short-wavelength photometry, where the excess emission due to accretion is more pronounced. As such, the Sloan $u$-band  can often be used as a proxy for accretion when no other tracer such as UV/optical spectra is available \citep{Gullbring1998, Fallscheer2006, Venuti2015, Guo2018, Flaischlen2022}. 

In some cases, the light curve morphology can relate to one of the several magnetospheric accretion regimes \citep{Romanova2008, Kurosawa2013, Blinova2016}. In the ``stable" regime, accretion occurs primarily through two, large accretion flows, typically present near each magnetic pole. These flows are consistent on the timescales of many rotations, though they do still exhibit some intrinsic accretion variability. Because of this stability, there is often associated periodic variability near the stellar rotation period as the accretion hotspot rotates with the star. Depending on viewing inclination, this can also cause the presence of red-shifted absorption in Balmer line profiles as the flow periodically obscures the view of the hotspot. In the ``unstable" regime, accretion occurs through many smaller, more transient accretion tongues, some of which can penetrate the magnetosphere near the equator, not just near the magnetic poles. These flows and associated hotspots can form and dissipate stochastically, often leading to light curves with little to no periodicity. 

The relationship between different photometric bands can inform us about the structure of the accretion hotspot. For example, \citet{Nature} measured a time lag between different bands in GM~Aur, where the short-wavelength $ug$ light curves peaked about 1 day before the long wavelength $ri$. They attributed this to an asymmetric, azimuthally elongated accretion hotspot on the surface of GM~Aur. The hotter, denser part of the hotspot (best traced by $ug$) rotated into/out of view prior to the cooler part of the hotspot (best traced by $ri$). \citet{Robinson2022} saw evidence for similar time  lags in a sample of 14 CTTSs. Such observations are consistent with 3D magnetospheric accretion models \citep{Romanova2004, Kulkarni2008}, which also predict accretion substructure and asymmetries. 

In order to better understand how time-variable accretion occurs and what effect it has on the star and disk, a large multi-epoch, multi-wavelength monitoring campaign was carried out for four CTTSs: TW~Hya, RU~Lup, BP~Tau, and GM~Aur. These observations covered several weeks in both 2021 and 2022 and include data products such as $HST$ UV spectra taken as part of the ULLYSES $HST$ Director's Discretionary Program and UV-NIR photometry and high resolution optical spectra obtained as part of the ODYSSEUS and PENELLOPE collaborations \citep[see][]{Espaillat2022, Manara2021}. Here, we present the results of the photometric monitoring as it relates to accretion, while \citet{PaperI} (hereafter Paper I) and \citet{PaperIII} (hereafter Paper III) focus on the UV and optical spectra, respectively. We first describe our observations and data in Section \ref{sec: Observations}. In Section \ref{sec: Analysis and Results} we present our light curves, the results of our periodicity analyses, color-magnitude diagrams, and search for correlations with accretion from results presented in Paper I. Next, in Section \ref{sec: Discussion}, we discuss these results in more detail and compare to previous work, including a search for time lags in the light curves and discussing the connection between photometric variability and accretion variability. We summarize our findings in Section \ref{sec: Conclusion}.

\section{Observations and Data Reduction} \label{sec: Observations}

Multi-wavelength, multi-epoch observations of the CTTSs TW~Hya, RU~Lup, BP~Tau, and GM~Aur were obtained in 2021 (Epoch 1/E1) and 2022 (Epoch 2/E2). UV-NIR photometric light curves are presented here, while contemporaneous $HST$ UV spectra and ground-based optical spectra are presented in Paper I and Paper III, respectively. More background information on the individual objects can be found in Paper I. These light curves were obtained with Las Cumbres Observatory Global Telescope (LCOGT), Konkoly Observatory, Crimean Astrophysical Observatory (CrAO), $Transiting Exoplanet Survey Satellite$ ($TESS$), American Association of Variable Star Observers (AAVSO), All-Sky Automated Survey for Supernovae (ASAS-SN), and Zwicky Transient Facility (ZTF). Details of these observations can be found in Table \ref{tab: Photometry}. The absolute flux calibration between the different sources of photometry varies, so we scale the various sources based on linear relationships between contemporaneous data. This includes scaling Johnson $RI$ to Sloan $ri$. See Appendix \ref{appendix: Scaling} for a description of this process. Below we present more details on the data obtained with each facility.

\begin{deluxetable*}{c c c c c c c}[htp]
\tablecaption{Summary of Photometric Observations \label{tab: Photometry}}
\centering
\tablehead{
\colhead{Object} & \colhead{Epoch} & \colhead{Source} & \colhead{Date (UT)} & \colhead{MJD} & \colhead{Total \#} & \colhead{Filter(s)} \\
\colhead{} & \colhead{} & \colhead{} & \colhead{[Begin/End]} & \colhead{[Begin/End]} & \colhead{of Points} & \colhead{}
}
\startdata
TW~Hya & 1 & LCOGT & 2020-12-16/2021-12-31 & 59199.3/59579.0 & 354 & $uVri$ \\
 &  & AAVSO & 2021-02-24/2021-08-25 & 59269.5/59451.0 & 894 & $BVRI$ \\ 
 &  & ASAS-SN & 2020-06-14/2022-01-19 & 59014.9/59598.1 & 229 & $g$ \\
 &  & \emph{TESS} & 2021-03-09/2021-04-01 & 59282.3/59305.5 & 2857 & \emph{TESS} \\
\hline
TW~Hya & 2 & LCOGT & 2022-03-20/2022-07-28 & 59658.2/59788.7 & 3873 & $ugVriz$ \\
 &  & AAVSO & 2022-03-03/2022-08-09 & 59641.1/59800.0 & 1048 & $BVRI$ \\
 &  & ASAS-SN & 2022-01-22/2023-02-20 & 59601.3/59995.2 & 216 & $g$ \\
\hline
\hline
RU~Lup & 1 & LCOGT & 2021-05-15/2021-10-04 & 59349.2/59491.0 & 192 & $uVri$  \\
 &  & AAVSO & 2021-04-26/2021-09-20 & 59330.2/59477.0 & 514 & $BVRI$ \\
 &  & ASAS-SN & 2020-06-08/2022-01-18 & 59008.8/59597.4 & 188 & $g$ \\
\hline
RU~Lup & 2 & LCOGT & 2022-02-01/2022-10-03 & 59611.4/59855.4 & 1374 & $uBgVriz$ \\
 &  & AAVSO & 2022-04-17/2022-10-18 & 59686.7/59870.0 & 387 & $BVRI$ \\
 &  & ASAS-SN & 2022-01-22/2023-02-20 & 59601.4/59995.3 & 165 & $g$ \\
\hline
\hline
BP~Tau & 1 & LCOGT & 2021-07-21/2021-12-13 & 59416.6/59561.3 & 1390 & $uBgVriz$ \\
 &  & AAVSO & 2020-07-15/2022-01-20 & 59045.4/59599.9 & 482 & $BVRI$ \\
 &  & ASAS-SN & 2020-07-13/2022-01-20 & 59043.6/59599.9 & 296 & $g$ \\
 &  & ZTF & 2020-08-03/2021-11-22 & 59064.5/59540.4 & 235 & $g$ \\
 &  & Konkoly & 2021-08-10/2021-09-12 & 59436.1/59469.9 & 68 & $BVri$ \\
 &  & \emph{TESS} & 2021-09-16/2021-11-05 & 59473.7/59523.9 & 6251 & \emph{TESS} \\
\hline
BP~Tau & 2 & LCOGT & 2022-02-28/2023-02-24 & 59638.2/59999.2 & 2277 & $uBgVriz$ \\
 &  & AAVSO & 2022-01-21/2023-02-09 & 59600.9/59984.3 & 865 & $BVRI$ \\
 &  & ASAS-SN & 2022-01-22/2023-02-21 & 59601.1/59996.2 & 250 & $g$ \\
 &  & ZTF & 2022-02-12/2022-11-06 & 59622.2/59889.3 & 72 & $g$ \\
 &  & Konkoly & 2022-11-25/2023-01-02 & 59908.2/59946.1 & 28 & $BVri$ \\
 &  & CrAO & 2022-08-12/2023-01-26 & 59803.0/59970.7 & 208 & $BVRI$ \\
 &  & \emph{TESS} & 2022-11-30/2022-12-23 & 59913.4/59936.2 & 7987 & \emph{TESS} \\
\hline
\hline
GM~Aur & 1 & LCOGT & 2021-07-29/2021-12-30 & 59424.2/59578.8 & 2095 & $ugVriz$ \\
 &  & AAVSO & 2020-07-27/2022-01-16 & 59057.4/59595.9 & 368 & $BVRI$ \\
 &  & ASAS-SN & 2020-07-20/2022-01-19 & 59050.6/59598.1 & 155 & $BVri$ \\
 &  & ZTF & 2020-08-03/2021-11-22 & 59064.5/59540.4 & 126 & $g$ \\
 &  & Konkoly & 2021-10-13/2021-12-14 & 59500.0/59562.8 & 136 & $BVri$ \\
 &  & \emph{TESS} & 2021-09-16/2021-11-05 & 59473.7/59523.9 & 6036 & \emph{TESS} \\
\hline
GM~Aur & 2 & LCOGT & 2022-02-24/2023-02-21 & 59634.1/59996.1 & 1590 & $ugVriz$ \\
 &  & AAVSO & 2022-01-21/2023-02-15 & 59600.9/59990.1 & 311 & $BVRI$ \\
 &  & ASAS-SN & 2022-01-22/2023-02-22 & 59601.4/59997.0 & 141 & $g$ \\
 &  & ZTF & 2022-02-13/2022-11-06 & 59623.2/59889.3 & 33 & $g$ \\
 &  & Konkoly & 2022-11-30/2022-12-13 & 59913.9/59926.7 & 16 & $BVri$ \\
 &  & CrAO & 2022-08-12/2023-01-26 & 59804.0/59970.7 & 180 & $BVRI$ \\
 &  & \emph{TESS} & 2022-11-28/2022-12-23 & 59911.6/59936.2 & 9395 & \emph{TESS} \\
\enddata
\tablenotetext{}{}
\end{deluxetable*}

\subsection{LCOGT Photometry} \label{sec: LCOGT Photometry}

$ugVriz$ photometry of all four targets was obtained in E1 and E2 with LCOGT's network of 0.4-, 1-, and 2-meter telescopes (PIDs: NSF2021B-015, NSF2022A-004, NSF2022B-019, CLN2021B-003, LCO2021B-001, LCOEP2020A-001, DDT2021A-001, DDT2021A-010, FTPEPO2014A-004, KEY2020B-009). Note that some LCOGT observations of GM~Aur from E1 (PID: CLN2021B-003, LCO2021B-001) were first presented in \citet{Bouvier2023}. 

BP~Tau and GM~Aur were observed with sub-nightly cadence in both E1 and E2 over a months-long baseline, while TW~Hya and RU~Lup were monitored with similar cadence and baseline only in E2. All four targets were observed with LCOGT close in time to their $HST$ observations. Some $uVi$ photometry was extracted and calibrated by ULYSSES \citep{ULYSSES}, but to ensure consistency, we perform our own photometric extraction and absolute flux calibration on all the LCOGT data. Our methods are as follows:

\begin{enumerate}
    \item Standard flat fielding and dark/bias subtraction are first performed by LCOGT's pipeline, BANZAI \citep{BANZAI}. We perform astrometric correction on each frame using Astrometry.net. Frames for which no solution could be found were discarded, except when an image was obtained within a set so that the WCS information from a nearby corrected image could be used.
    \item We select all sources in the frame cross-matched with the ATLAS-REFCAT 2 \citep{ATLAS-REFCAT2} survey, down to about 16th magnitude in $g$. We then perform initial aperture photometry using a 20-pixel wide aperture and a background annulus between 30 and 40 pixels. Sources with SNR$<$5 were discarded.
    \item We fit each source assuming the product of 2D Gaussian and Moffat profiles using least squares. The flux of each source is taken to be the total integral of the resulting 2D profile, while the uncertainty is taken from a background annulus between 30 and 40 pixels. The pixel scale varies between 0$\arcsec$.58--0$\arcsec$.78.
    \item We calculate a magnitude zero-point for each source using the fitted flux and catalog magnitude. For $griz$, we use the ATLAS-REFCAT 2 \citep{ATLAS-REFCAT2} catalog, which yields the most cross-matched sources and the tightest correlations. To calibrate $u$, we use the Guide Star Catalog \citep{GSC}, version 2.4.2.
    \item Magnitude zero-points were converted to flux zero points. 3$\sigma$ outlying flux zero-points were discarded and the average flux zero-point was then converted back to an average magnitude zero-point. This was used to calculate a calibrated apparent magnitude for each source in the frame, including the target. 
\end{enumerate}

\subsection{Konkoly Observatory Photometry} \label{sec: Konkoly Observatory Photometry}

We observed BP~Tau in E1 between August 9 and September 12, 2021 (MJD: 59436.6--59470.4) and GM~Aur in E1 between October 13 and December 15 2021 (MJD: 59500.5--59563.3), contemporaneously with $HST$, with a roughly nightly cadence using the RC80 telescope of the Piszk\'{e}stet\H{o} Mountain Station of Konkoly Observatory (Hungary). We employed a set of Bessel $BV$ and Sloan $ri$ filters and took three images per night with each filter. After the usual bias, dark, and flat--field corrections, we computed aperture photometry for the target and a set of comparison stars using an aperture radius of 5 pixels and a sky annulus between 20 and 40 pixels (the pixel scale is 0$\farcs$55). 

For the differential photometry, we used a set of 57/50 comparison stars in the 18$'\times$18$'$ field of view and fit a linear color term for BP~Tau/GM~Aur, respectively. The conversion to the standard system was done using the $BVri$ magnitudes of the comparison stars from the APASS9 catalog \citep{APASS}. The final uncertainties are the quadratic sum of the formal uncertainty of the aperture photometry, the uncertainty of the photometric calibration, and the scatter of the target magnitudes measured on the same night with the same filter from three observations.

\subsection{CrAO Photometry} \label{sec: CrAO Photometry}

$BVRI$ photometry of BP~Tau and GM~Aur was obtained in E2 with the Crimean Astrophysical Observatory (CrAO) 1.25-m AZT-11 telescope from August 12, 2022 to January 27, 2023 (MJD: 59803.5--59971.2). We used a Greateyes GE 2048x2048 BI MID CCD camera with 13.5 $\mu$m pixels and a two-stage cooling. Exposure times per image were 180, 60, 30, and 15 s in $BVRI$, respectively. After standard reduction steps on bias, dark, and flat--field, we extracted aperture photometry. The results from five separate images in each filter obtained during the same night were averaged, resulting in typical internal uncertainties of 0.008–0.010 mag in $B$, 0.004–0.005 mag in the $VRcIc$.
 
We performed differential photometry between BP~Tau and a non-variable, nearby comparison star, HD~281930, whose brightness and colors are $V$ = 11.351, ($B-V$) = 0.981, ($V-R$) = 0.556 and ($V-I$) = 1.069. A nearby control star of similar brightness, 2MASS~04190416+2912331, was used to verify that the comparison star was not variable. Similarly, in the case of GM~Aur, we used 2MASS~J04551015+3021333 as the comparison star, whose brightness and colors are $V$ = 11.834, ($B-V$) = 1.419, ($V-R$) = 0.749 and ($V-I$) = 1.401. As a control star, we used HD~282626. We convert the Johnson-Cousins $RI$ photometry to Sloan $ri$ using a simple linear correction term.

\subsection{\emph{TESS}} \label{sec: TESS Photometry}

We utilize data from \emph{TESS} for TW~Hya in E1, BP~Tau in E1 and E2, and GM~Aur in E1 and E2. \emph{TESS} observed TW~Hya for 23 nights in Sector 36 from March 9 to April 4, 2021 (MJD: 59282.8--59306.0), covering about 6 full rotations. BP~Tau and GM~Aur were observed simultaneously in Sectors 43 and 44 for 48 nights from September 17 to November 6, 2021 (MJD: 59474.2--59524.4), then again for 24 nights in Sector 59 from November 29 to December 23, 2022 (MJD: 59912.1--59936.7). We use the $TESS$-$Gaia$ Light Curve \citep[{\it tglc};][]{tglc} package to obtain reduced light curves for each target in each sector, then scale the resulting fluxes to simultaneous $i$ band light curves.

\subsection{AAVSO, ASAS-N, and ZTF Photometry} \label{sec: Other Photometry}

We supplement our data with $BVri$ photometry from the AAVSO\footnote{https://www.aavso.org} for all 4 targets, $g$ photometry from ASAS-SN \citep{ASAS-SN1, ASAS-SN2} for all 4 targets, and $g$ photometry from the ZTF \citep{ZTF1} for BP~Tau and GM~Aur.

AAVSO observations are obtained by dozens of different observers and in various conditions and are thus subject to variations in calibration, resulting in inconsistencies/artifacts in the light curves. In an attempt to ameliorate this, we discard data from observers with fewer than 25 total observations across all bands. This resulted in a balance of quality and quantity of observations, leaving primarily observers with many observations and more consistent calibration and cadence. Some AAVSO observations are obtained in clusters with minutes-long cadence and can produce artifacts and aliasing in our periodograms (see Section \ref{sec: Analysis and Results}). We thus perform 30-minute median-binning of the AAVSO light curves, where a bin's uncertainty is the quadratic sum of standard deviation in that bin and the mean photometric uncertainty. Note that our analyses in Section \ref{sec: Analysis and Results} were also performed without the AAVSO data and generally resulted in similar conclusions at lower significance. ASAS-SN and ZTF data are self-consistently reduced and calibrated and did not require any pruning.

\section{Analysis and Results} \label{sec: Analysis and Results}

\begin{figure*}[ht]
    \centering
    \includegraphics[width=0.905\textwidth]{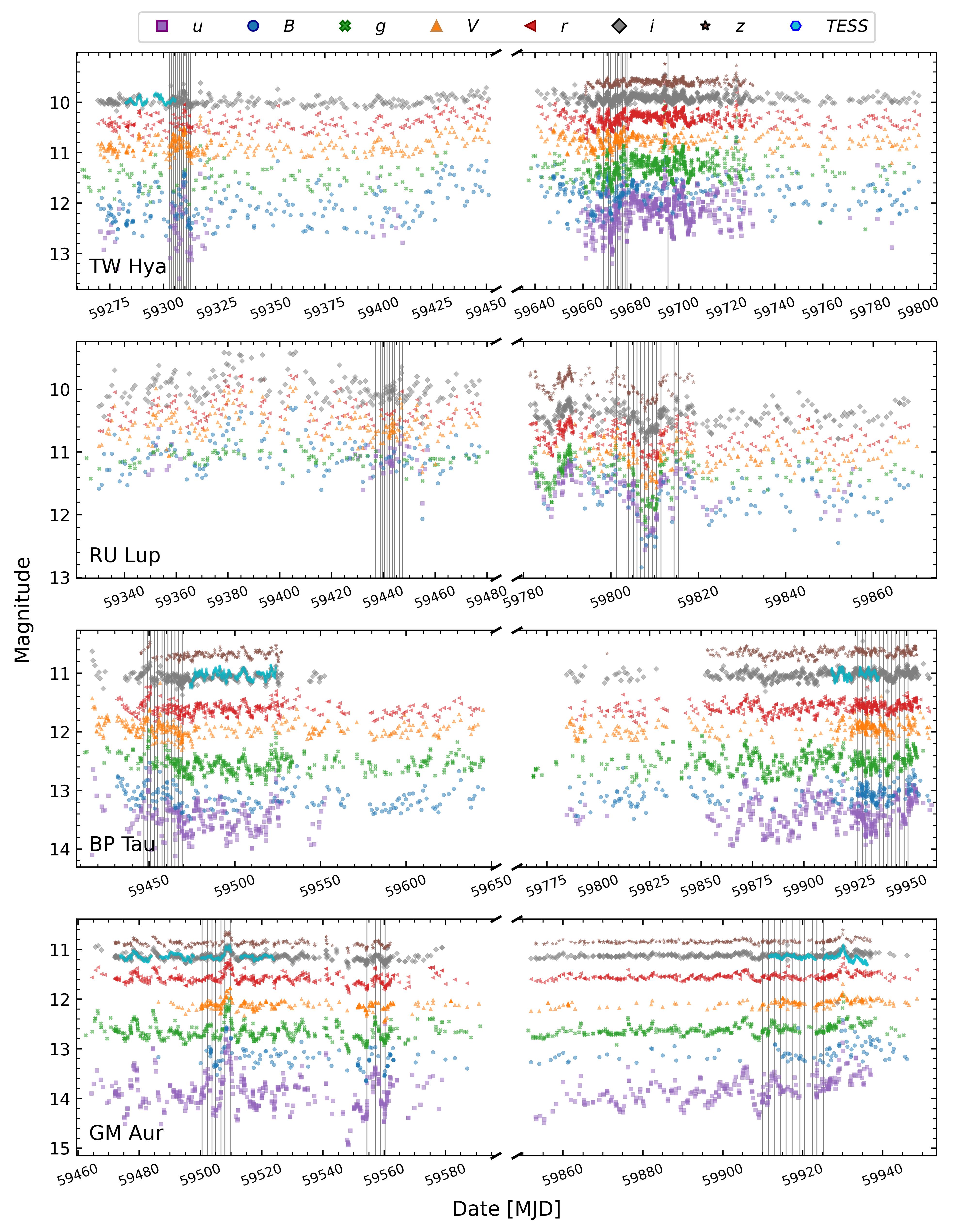}
    \caption{Optical light curves for TW~Hya, RU~Lup, BP~Tau, and GM~Aur (top to bottom). Symbols corresponding to $uBgVriz/TESS$ data are labeled in the key. $TESS$ has been scaled to $i$. Error bars are shown, but are often smaller than the symbol. The solid grey vertical lines correspond to the times of the $HST$ observations. More details on the source of the photometry is given in Table \ref{tab: Photometry}.}
    \label{fig: Light Curves}
\end{figure*}

\begin{figure*}[ht]
    \centering
    \includegraphics[width=0.95\textwidth]{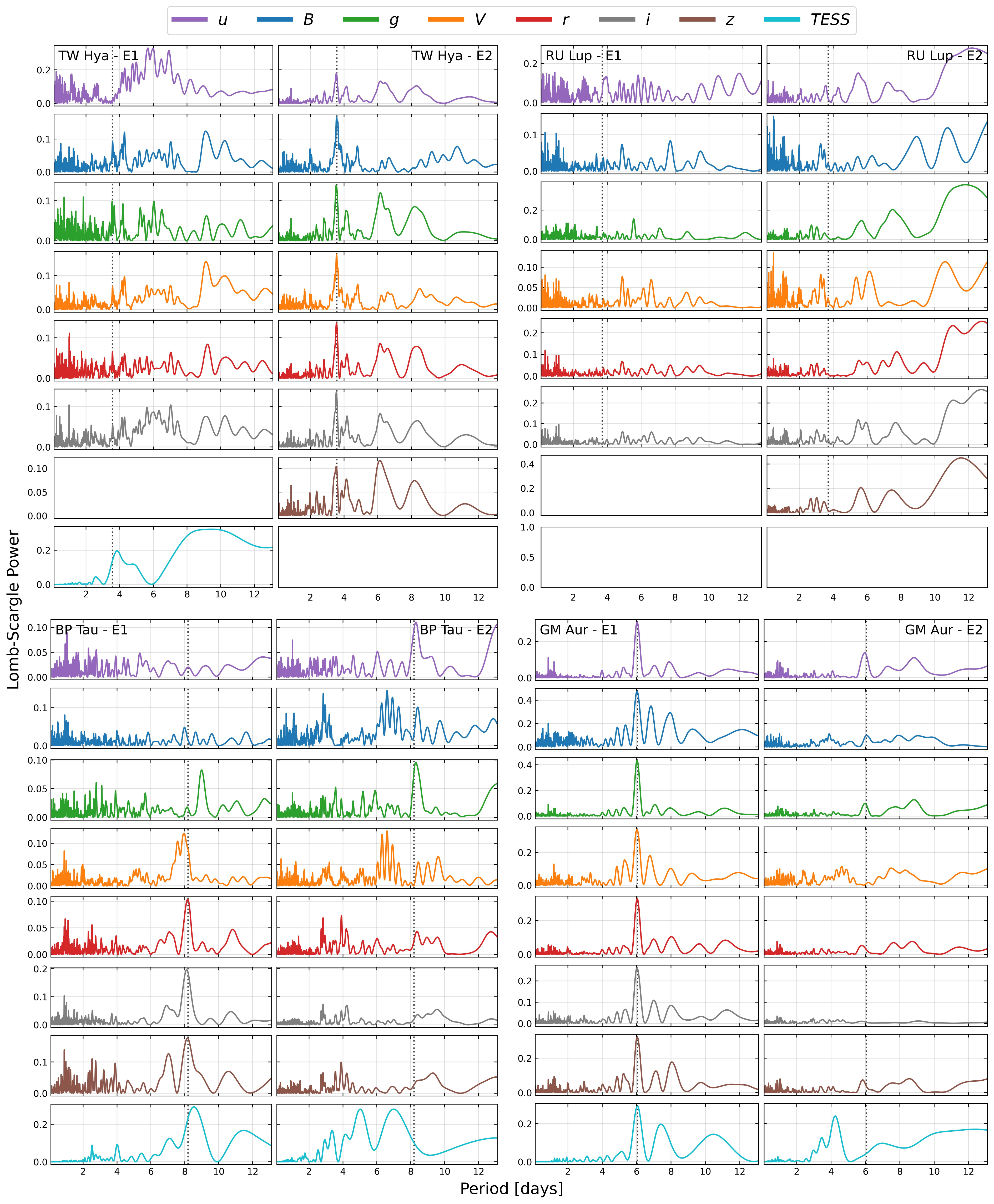}
    \caption{Lomb-Scargle periodograms for TW~Hya (top left), RU~Lup (top right), BP~Tau (bottom left), and GM~Aur (bottom right) after subtracting a linear fit. Dashed black line is the assumed rotation period of the star from previous studies (see Table 1, Paper I), not necessarily those calculated here.} 
    \label{fig: Light Curve Periods}
\end{figure*}

\begin{figure*}[ht]
    \centering
    \includegraphics[width=0.99\textwidth]{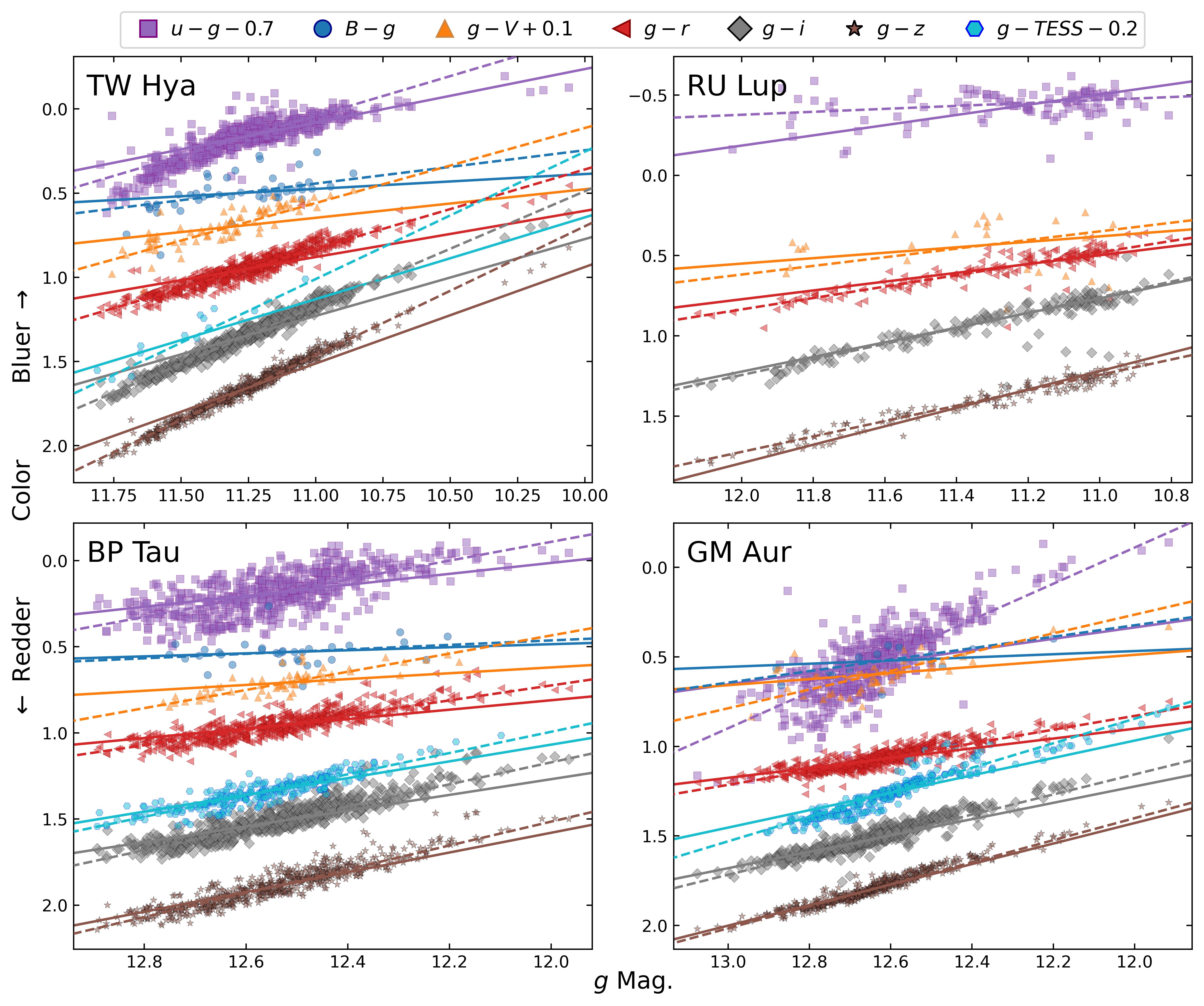}
    \caption{Color-magnitude diagrams for each target. Solid lines represent the slope of the reddening curve from \citep{Cardelli1989} and dashed lines are a linear RANSAC regression fit to each set. Note that offsets are applied to some colors for visual distinction, as noted in figure legend.}
    \label{fig: Light Curve Colors}
\end{figure*}

Figure \ref{fig: Light Curves} shows the $uBgVriz/TESS$ light curves for TW~Hya, RU~Lup, BP~Tau, and GM~Aur (top to bottom) in both E1 and E2. Here we quantify the light curve properties such as periodicity, timescales of variability, and color variability.  We then compare our results to those from Paper I to search for correlations with accretion and measure $Q$ and $M$ variability metrics following \citet{Cody2014}. Figure \ref{fig: Light Curve Periods} shows Lomb-Scargle periodograms for each band and target. We subtract a linear fit from each lightcurve before calculating the periodogram. Figure \ref{fig: Light Curve Colors} shows a color-magnitude diagram for each target and light curve, along with a comparison to the slope of variable extinction. 

All 4 targets show considerable variability at short wavelengths ($\ge$1 mag in $uBg$), generally with smaller amplitude at longer wavelengths ($\le$0.5 mag in $riz$). Most of the variability is on timescales of a few days, but some notable variability on shorter timescales is seen. Variability on longer timescales (such as between E1 and E2) is seen, both in baseline brightness and in light curve behavior. Below we discuss each target and light curve in more detail and search for correlations with accretion. 

\subsection{Light Curves of TW~Hya}

TW~Hya shows classic burster-like behavior \citep[i.e., deviations from quiescent brightness tend to be positive; see Section \ref{sec: QM Results} and][]{Cody2014} with moderate periodicity (Figure~\ref{fig: Light Curves}). It appears to be approximately equally bright throughout E1 and E2. Most of the variability lasts no more than about 2 rotation periods \citep[6--10 days;][]{Siwak2018}. One potential exception may be a period of brightening ($\Delta u\sim$0.5 mag) from MJD$\sim$59410--59440 towards the end of E1. This is reflected in all bands and is superimposed on TW~Hya's shorter-term periodic behavior. Another exception may be a roughly 20-day period of dimming near MJD=59670 that is also superimposed on a periodic signature.

Figure \ref{fig: Light Curve Periods} shows that TW~Hya exhibits some periodic signals near its assumed 3.57-day rotation period in E1, but often at insignificant levels. A $\sim$4.27-day period dominates $BV$ and is seen in other bands though below the 0.1\% significance level. Our $TESS$ light curves shows a peak at 3.85 days, as well as a broad signal from $\sim$8--12+ days. The 3.84-day period is likely related to the rotation of TW~Hya. The broad signal near 10 days may be related to a $\sim$9.1-day period seen in $BVri$, but should be considered carefully; it approaches half the $TESS$ baseline and is not seen after removing the 3.57-day periodic signal (see Appendix \ref{appendix: Periodicity}).

In E2, we robustly detect a 3.54-day period in every band ($uBgVriz$) with high power and contrast. The 4.27-day period from E1 may be weakly detected in all bands. Also seen in $ugriz$ are broad signals near 6.1 and 8.2 days. The periodicities near 3.54 and 4.26 days in E1 and E2 are roughly in line with previous studies that found a range of periodicities between 1.4--4.7 days \citep{Huelamo2008, Siwak2011, Siwak2014, Siwak2018, Herczeg2023, SiciliaAguilar2023}, with 3.57 days being the assumed rotational period for TW~Hya. While some statistically significant, jagged peaks are seen between 1.0--3.0 days, they are not consistent between bands or between epochs, and should not be considered physical.

\subsection{Light Curves of RU~Lup}

RU~Lup appears dimmer in E2 ($\Delta i\sim$0.5, $\Delta B\sim$1.0 mag), consistent with the change in accretion rate between E1 and E2 presented in Paper I (Figure~\ref{fig: Light Curves}). Its variability is mostly symmetric, though it underwent at least one strong dimming event in E2 (MJD$\sim$59810) during the $HST$ observations. RU~Lup also shows comparable variability amplitudes at both long ($riz$) and short wavelengths ($ug$), in contrast with the other targets that show lower amplitude variability in $riz$. 

In E1, RU~Lup does not show significant periodicity in any band. In E2, broad signals are detected at significance near 5.5--5.7 days in $ugVri$ and 7.5--7.8 in $ugri$, while no peaks are seen in $B$. Given their wide breadth, low significance, and inconsistency amongst the 6 filters, we do not expect that these periodicities correspond to any dominant physical, periodic process in RU~Lup. No periodicity, even insignificant, is detected at or near the assumed 3.71-day period \citep{Stempels2007}. This is consistent with previous photometric studies of RU~Lup \citep{Giovannelli1995, Percy2010, Siwak2016}, many of which did not recover any periodicity.

\subsection{Light Curves of BP~Tau}

BP~Tau's light curves appear to be mostly symmetric, but with some short, sporadic dips and bursts (Figure~\ref{fig: Light Curves}). It is marginally brighter in E2 than in E1 and does not appear to exhibit any long-term (on the scale of 1+ months) trends.

In E1, no periodicity is seen in $uB$. A strong power peak at 8.9-days is seen in $g$, though no other filter in either E1 or E2 shows a corresponding period. In $Vriz$ we see signal near 7.95--8.15 days, and 8.54 days in $TESS$. $TESS$ also shows two sharp peaks at 2.53 and 3.98 days, and two broad peaks at 7.03 and 11.20 days. With the exception of the 7.03-day signal (which may be seen in $riz$, albeit slightly shorter and at low significance), none of the $TESS$ peaks are seen in other filters. The $\sim$8.15-day period in $riz$ is consistent with the assumed 8.19-day period in BP~Tau \citep{Percy2006a}, and the 7.94-period in $V$ likely corresponds to this signal. 

In E2, $ug$ show clear periodicity at 8.32 days and $B$ at 8.4 days. Unlike in E1, $Vriz$ do not show the rotational period near 8.15-days, instead exhibiting jagged, sporadic peaks between 2.84--4.24 days. Like E1, detected periods in $TESS$ (3.36, 5.02, 7.00 days) do not correspond to obvious physical processes in BP~Tau (like rotation) or signals seen in $uBgVriz$. In neither epoch do we see evidence pointing towards the previously detected 7.6-day period \citep{Vrba1986, Simon1990, Osterloh1996}.

\subsection{Light Curves of GM~Aur}

GM~Aur exhibits significant variability in E1, exhibiting periodic-burster \citep[i.e. stochastic brightening coincident with an overall periodic behavior; see][]{Cody2014} type behavior (Figure~\ref{fig: Light Curves}). A moderate accretion burst ($\Delta u \sim$1.5 mag) near MJD=59509 is recovered well here (along with several other smaller bursts) and appears to last from MJD=59507--59511. Another moderate burst is seen in E2 near MJD=59930. In general, GM~Aur is equally as bright in E1 and E2, but shows vastly diminished variability in E2, either periodic or stochastic. The light curve also appears to rise gradually in E2, by about 0.5 mag in $u$, to $<$0.1 in $riz$. 

The known rotation period of GM~Aur of about 6 days \citep{Percy2006b, Nature, Bouvier2023} is detected in every filter ($uBgVriz/TESS$) in E1 at about 6.01 days. Each filter (except $TESS$) also shows peaks near 7 and 8 days, though with lower significance and with scatter of up to 0.20 days. $uBgVr$ exhibit a sharp but significant peak at 0.86 days. $TESS$ shows a large peak at 7.42 days not seen in the other filters. In E2, the 6-day rotation period persists, though with greatly diminished significance. It is detected only in the bluer bandpasses ($ug$) but is not robustly detected in any other filter ($BVriz/TESS$). $TESS$ exhibits a notable period at 4.26 days which is not seen with any significance in any other band.

\subsection{Correlations with Accretion} \label{sec: Correlations with Accretion}

\begin{figure*}
    \centering
    \includegraphics[width=0.97\textwidth]{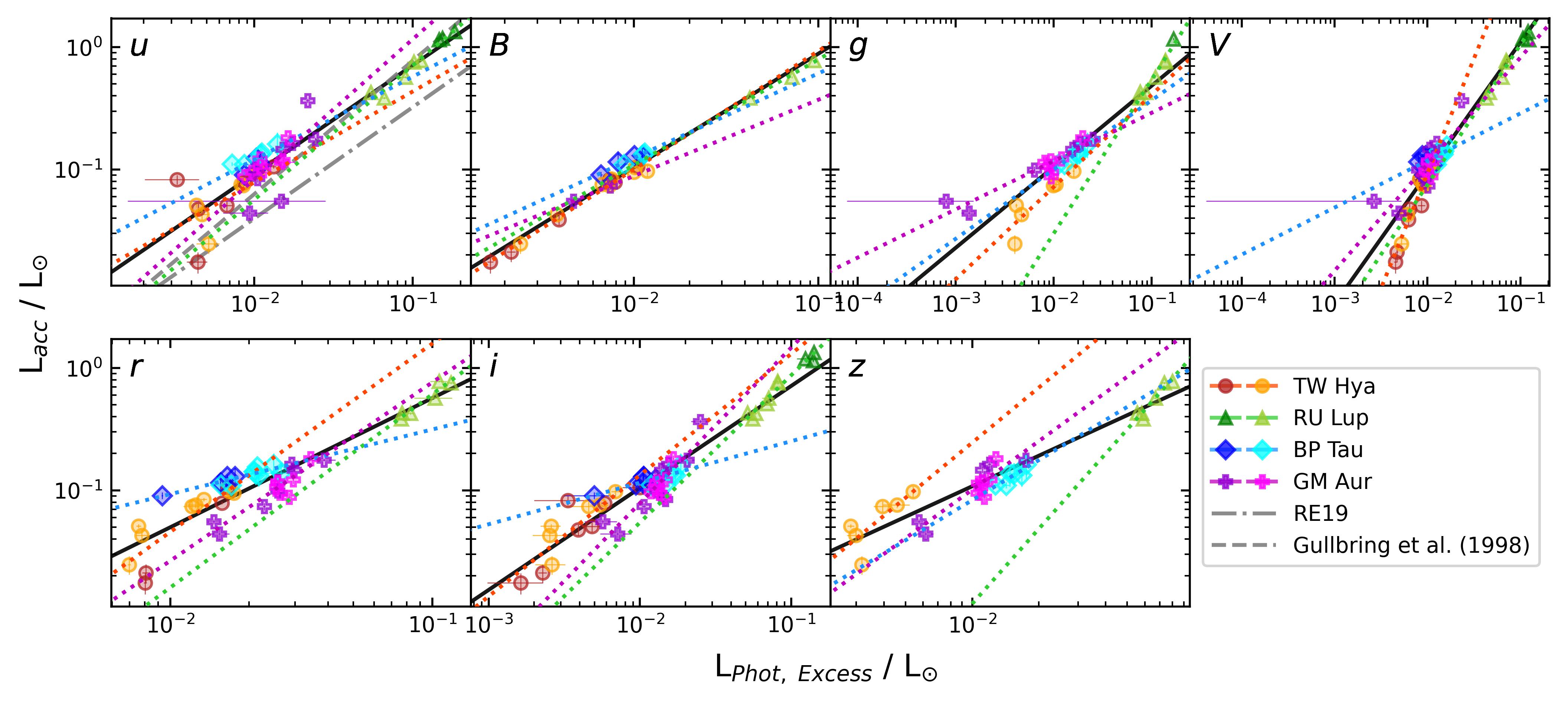}
    \caption{\lacc vs \lphotexcess in the $uBgVriz$ bands for all four targets: TW~Hya (red/orange), RU~Lup (dark/light green), BP~Tau (blue/cyan), and GM~Aur (purple/pink). Darker (filled circles)/lighter (open squares) points are observations from E1/E2, respectively. Solid, colored lines are the log-log linear fits to each individual target. Solid black line is the fit to the entire sample, dotted grey line is the $u\prime$ relationship from \citet{RE19}, and the grey dash-dotted line is the $U$ relationship from \citet{Gullbring1998}.}
    \label{fig: Lacc vs Photometry}
\end{figure*}

Here we attempt to connect our photometric data to the accretion rates obtained contemporaneously using $HST$ data in Paper I. Below we assume that all the photometric variability is related to accretion variability and that there is no significant variability within 2 hours. Both of these are unlikely to be the case all of the time.  However, in the interest of connecting these unprecedented contemporaneous photometric and UV spectroscopic datasets, we proceed while noting the above caveats. Figure \ref{fig: Lacc vs Photometry} shows the accretion luminosity \lacc vs excess photometric luminosity \lphotexcess for the 7 filters we utilize. \lacc is calculated using Equation \ref{eq: Lacc} with the accretion rates (\mdot) reported in Paper I, calculated using the accretion shock models from \citet{Gullbring1998} and \citet{RE19}. 

\begin{equation} \label{eq: Lacc}
    L_{acc} = \frac{GM_{\star}\dot{M}}{R_{\star}}(1-\frac{1}{R_i})
\end{equation}

To calculate \lphotexcess, we first convert our apparent magnitudes to total normalized luminosities in the given bandpass via 
\begin{equation} \label{eq: m to L}
    L_{i} = 10^{-m_i/2.5}\cdot ZP_i \cdot (c/\lambda_{eff}) \cdot W_{eff, i} \cdot 4\pi d^2/L_{\star}
\end{equation}
where $m$, $ZP$, $\lambda_{eff}$, and $W_{eff}$ are the apparent magnitude, flux zero-point, effective wavelength, effective width of the given filter $i$ \footnote{Zero-points, effective wavelengths, and effective widths from svo2.cab.inta-csic.es/theory/fps/index.php?gname=LasCumbres for filters on LCOGT telescopes}, \mstar is the stellar mass, \rstar is the stellar radius, \rin is the inner disk radius, 5\rstar, \dstar is the distance, \lstar is the stellar luminosity, and $c$ is the speed of light. See Paper I, Table 2 for these stellar parameters. 
Note that this conversion is only an approximation of the true total luminosity as measured by some filter $i$. Provided the underlying spectrum is not strongly convex/concave around $\lambda_{eff}$ (which they are generally not in our case), this approximation should be suitable.

From here, we estimate the underlying, non-accreting photospheric emission using WTTS spectra from \citet{Manara2013, Stelzer2013} of TWA 6 (for TW~Hya, RU~Lup, and BP~Tau) and TWA 9A (for GM~Aur). These are scaled to the photospheric levels estimated in Paper I using optical veiling measurements from Paper III. Then for each observation we subtract the non-accreting photospheric contribution at the given bandpass' effective wavelength. Our final \lphotexcess is the weighted average of all points within 2 hours of each $HST$ visit in Paper I, where data closer in time to the $HST$ observation are weighted more heavily. The uncertainty in \lphotexcess is the standard deviation of those points, though in cases with only one contemporaneous photometry point, we assume a 20\% uncertainty.

We fit log-log linear relationships ($log_{10}(L_{acc})$ vs $log_{10}(L_{\mathrm{Phot,\ Excess}})$) to the entire sample (solid black line, Fig.~\ref{fig: Lacc vs Photometry}, as well as to each target (solid colored lines). Globally, we find strong correlations (see Table \ref{tab: Lacc vs Lphotexcess Coefficients}) for all bands. We find a slightly more shallow relationship between \lacc and L$_{u/U,\ Excess}$ than \citet{Gullbring1996, RE19} with enhanced \lacc. Our higher \lacc was noted in Paper I, likely the result of our multi-column model. The trends of individual targets differ in some cases. TW~Hya and GM~Aur deviate from the global trend most strongly, especially in $uVi$, and exhibit the largest scatter. RU~Lup deviates slightly in some bands, generally with high \lacc/low \lphotexcess. \lacc in BP~Tau shows more shallow relationships to \lphotexcess than the other targets and the global relationship, similar to \lacc vs L$_{UV}$ in Paper I.

\begin{deluxetable}{c c c c}[ht]
    \tablewidth{100pt}
    \tablecaption{\lacc vs. \lphotexcess Log-log Linear Fit Coefficients \label{tab: Lacc vs Lphotexcess Coefficients}}
    \centering
    \tablehead{
    \colhead{$\ \ \ \ $Filter$\ \ \ \ $} & \colhead{$\ \ \ \ \ \ m\ \ \ \ \ \ $} & \colhead{$\ \ \ \ \ \ b\ \ \ \ \ \ $} & \colhead{$\ \ \ \ \ \ r\ \ \ \ \ \ $} 
    }
\startdata
$u$ & $0.88^{+0.02}_{-0.03}$ & $0.73^{+0.04}_{-0.06}$ & $0.93^{+0.01}_{-0.02}$ \\
$B$ & $0.93^{+0.03}_{-0.02}$ & $0.88^{+0.05}_{-0.04}$ & $0.980^{+0.004}_{-0.005}$ \\
$g$ & $0.66^{+0.04}_{-0.06}$ & $0.34^{+0.07}_{-0.12}$ & $0.92^{+0.02}_{-0.04}$ \\
$V$ & $1.07^{+0.04}_{-0.12}$ & $1.10^{+0.07}_{-0.21}$ & $0.94^{+0.01}_{-0.04}$ \\
$r$ & $1.05^{+0.02}_{-0.02}$ & $0.80^{+0.04}_{-0.03}$ & $0.911^{+0.007}_{-0.009}$ \\
$i$ & $0.86^{+0.02}_{-0.05}$ & $0.73^{+0.04}_{-0.08}$ & $0.951^{+0.009}_{-0.017}$ \\
$z$ & $0.83^{+0.02}_{-0.02}$ & $0.69^{+0.03}_{-0.03}$ & $0.946^{+0.009}_{-0.008}$ \\
\enddata
\tablecomments{Log-log linear fits are of the form: $log_{10}(\frac{L_{acc}}{L_{\odot}})=m*log_{10}(\frac{L_{\mathrm{Phot,\ Excess}}}{L_{\odot}})+b$. $r$ is the Pearson correlation coefficient.}
\end{deluxetable}

\begin{figure*}
    \centering
    \includegraphics[width=0.975\textwidth]{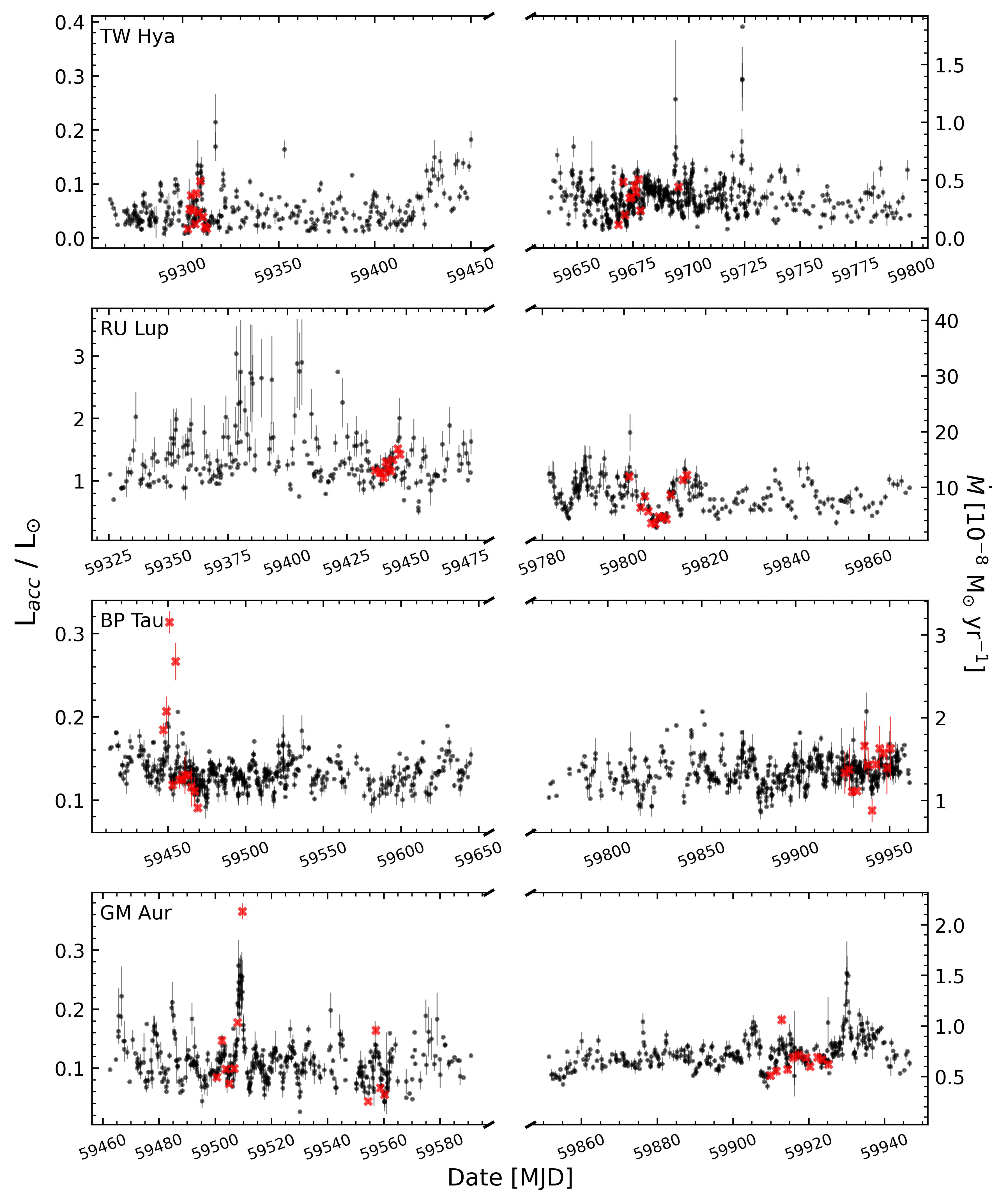}
    \caption{\lacc (left axis) and \mdot (right axis) over time for all 4 targets derived from relationships pictured in Figure \ref{fig: Lacc vs Photometry}. Red crosses are accretion luminosities derived in Paper I from shock modeling.}
    \label{fig: Lacc Curves}
\end{figure*}

From the relationships between \lacc and \lphotexcess derived above, we can attempt to see how \lacc (and by extension \mdot) changes over time by creating plots of \lacc and \mdot over time. To do so, we convert \lphotexcess to \lacc using the fit coefficients for each filter and target. Some poorly-fit filters or those with few simultaneous points are ignored, including $u$ for TW~Hya, BP~Tau, and GM~Aur and $B$ for GM~Aur. We then combine the resulting \lacc and \mdot and bin to 2-hour segments. The resulting \lacc and \mdot are shown in Figure \ref{fig: Lacc Curves} along with the \lacc and \mdot derived in Paper I. While the connection between bluer filters ($uBg$) and accretion is generally stronger \citep{Ingleby2013, Robinson2022}, we elect to utilize all filters to construct the \lacc and \mdot. Not only does this improve the sampling and SNR of \lacc and \mdot, but using only the bluest bands does not appreciably improve the correspondence between the \lacc derived in Paper I and \lacc and \mdot over time. We discuss Figure \ref{fig: Lacc Curves} further in Section \ref{sec: Discussion}.

\subsection{Q and M Variability Metrics} \label{sec: QM Results}

To further classify our light curves, we utilize the $Q$ and $M$ variability metrics originally developed in \citet{Cody2014}. $Q$ is a measure of the light curve's periodicity (or lack thereof). It generally takes values between 0 and 1, with 0 being purely periodic and 1 being purely stochastic. $M$ is a measure of the asymmetry in a light curve, whether deviations from the mean tend to be positive (bursts) or negative (dips). It generally takes values between -1 and 1, with lower values representing bursters, higher values being dippers, and values near 0 representing symmetric light curves. To calculate $Q$ and $M$, we follow the approaches of \citet{Cody2018} and \cite{Robinson2022}.

For $Q$, unlike \citet{Cody2018} and \citet{Robinson2022}, we do not calculate the period in the same way; we use the periods determined solely from our Lomb-Scargle analyses or from the literature when no period could be found. These are 3.54, 3.71, 8.15/8.31, and 6.01 days for TW~Hya, RU~Lup, BP~Tau, and GM~Aur, respectively. We use two periods for the corresponding epochs in BP~Tau, as the detected period differs in each epoch. In practice, small differences of up to 0.2 days in the adopted periods made little difference in the final value of $Q$. Using these periods, we fold the light curve and fit a Gaussian Process (GP) to three horizontally stacked copies of the phase-folded light curve. Here we use a length-scale of 0.3, about 1/3 of a full period. The fit to the central portion of this stacked light curve is extended and subtracted from the full, raw light curve. From here, we calculate $Q$ as:
\begin{equation}
    Q = \frac{\sigma_{resid}^2 - \sigma_{phot}^2}{\sigma_m^2 - \sigma_{phot}^2}
\end{equation}
where $\sigma_{resid}$ is the standard deviation of the residuals of the GP-subtracted light curve, $\sigma_m$ is the standard deviation of the raw light curve, and $\sigma_{phot}$ is 1.25 times the median photometric uncertainty. 

\begin{figure}
    \centering
    \includegraphics[width=0.47\textwidth]{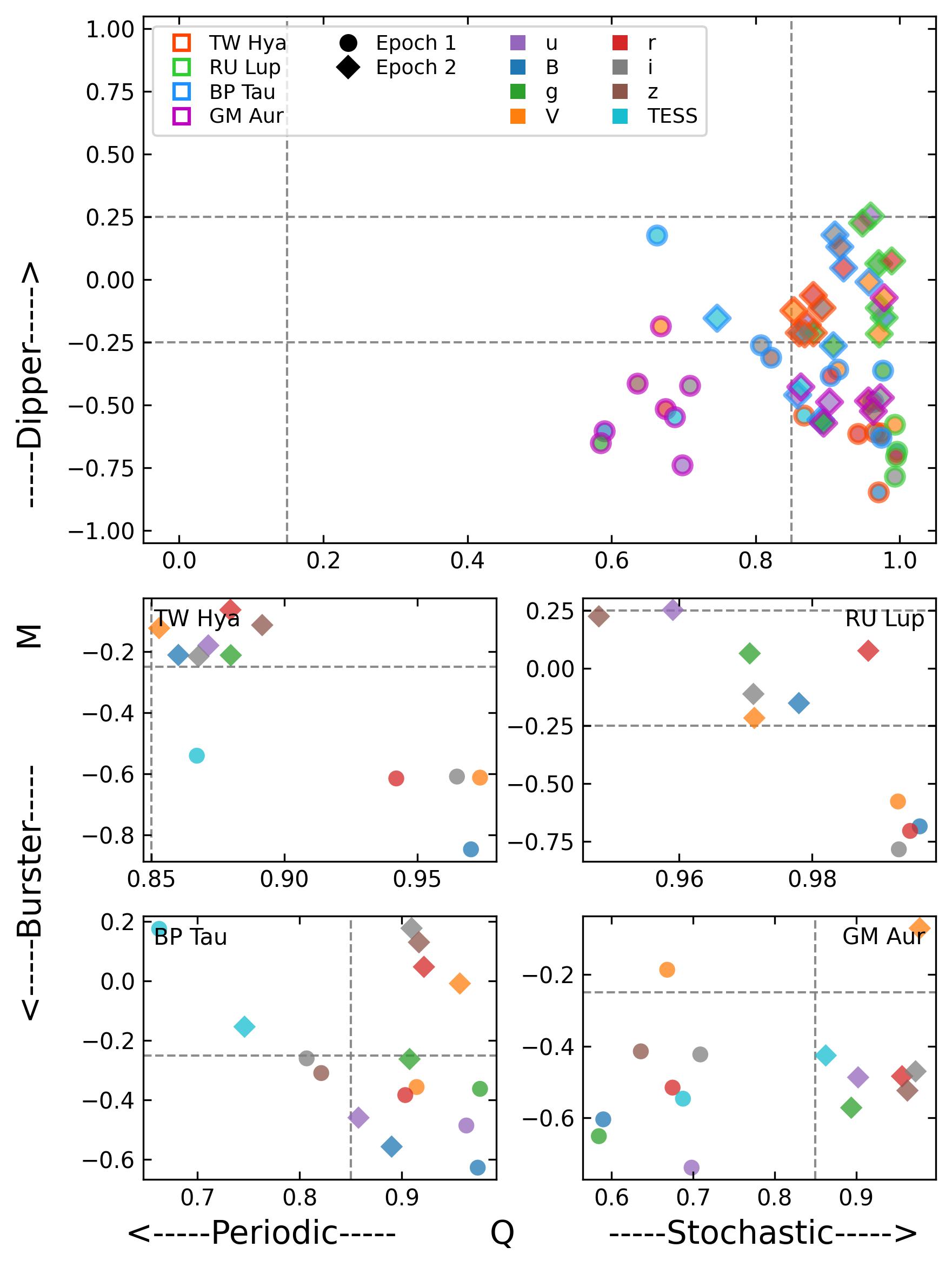} 
    \caption{$M$ vs. $Q$ for all light curves in our sample. Top panel is all targets together, while bottom panels show each target individually for clarity. Marker edge color denotes the object, with red, green, blue, and purple being TW~Hya, RU~Lup, BP~Tau, and GM~Aur, respectively. Marker fill color denotes the bandpass. Circles and diamonds are Epoch 1 and Epoch 2, respectively. Grey dashed lines delineate the various variability classes laid out by \citet{Cody2018}.}
    \label{fig: QM}
\end{figure}

For $M$, we first determine a smoothed version of the raw light curve using GP. This method is employed by \citet{Robinson2022}, though our method differs in that we use a length scale of 12 hours instead of 2, to account for the inferior sampling and cadence of most of our light curves. We then subtract this smoothed curve from the raw light curve and remove any 5-$\sigma$ outlying points. Finally, we calculate $M$ as
\begin{equation}
    M = -\frac{\langle d_{10\%} \rangle - d_{med}}{\sigma_d}
\end{equation}
where $\langle d_{10\%} \rangle$ is the average of the top and bottom 10\% of the clipped light curve, d$_{med}$ is the median of the clipped light curve, and $\sigma_d$ is the standard deviation of the clipped light curve. Our results for both $Q$ and $M$ are shown in Figure \ref{fig: QM} and Table \ref{tab: QM}. 

\begin{deluxetable*}{c | c c | c c | c c | c c}[ht]
\tablecaption{$Q$ \& $M$ for Each Light Curve in Our Sample} \label{tab: QM}
\centering
\tablehead{
    \multicolumn{1}{c|}{Filter} & \multicolumn{2}{c|}{TW~Hya} & \multicolumn{2}{c|}{RU~Lup} & \multicolumn{2}{c|}{BP~Tau} & \multicolumn{2}{c}{GM~Aur} \\
    \multicolumn{1}{c|}{} & \multicolumn{1}{c}{$Q$} & \multicolumn{1}{c|}{$M$} & \multicolumn{1}{c}{$Q$} & \multicolumn{1}{c|}{$M$} & \multicolumn{1}{c}{$Q$} & \multicolumn{1}{c|}{$M$} & \multicolumn{1}{c}{$Q$} & \multicolumn{1}{c}{$M$} 
}
\startdata
$u$ & \ \ *\ \ /0.87 & \ \ \ *\ \ /-0.18 & \ \ *\ \ /0.95 & \ \ \ *\ \ /\ 0.25 & 0.96/0.86 & -0.49/-0.46 & 0.70/0.90 & -0.74/-0.49 \\
$B$ & 0.97/0.86 & -0.85/-0.21 & 1.00/0.98 & -0.69/-0.15 & 0.97/0.89 & -0.63/-0.56 & 0.59/\ \ *\ \ \  & -0.60/\ \ *\ \ \ \ \\
$g$ & \ \ *\ \ /0.88 & \ \ \ *\ \ /-0.21 & \ \ *\ \ /0.97 & \ \ \ *\ \ /\ 0.06 & 0.98/0.91 & -0.36/-0.26 & 0.58/0.89 & -0.65/-0.57 \\
$V$ & 0.97/0.85 & -0.61/-0.12 & 0.99/0.97 & -0.58/-0.22 & 0.91/0.96 & -0.35/-0.01 & 0.67/0.98 & -0.19/-0.07 \\
$r$ & 0.94/0.88 & -0.61/-0.06 & 0.99/0.99 & -0.71/0.08 & 0.90/0.92 & -0.38/0.05 & 0.67/0.96 & -0.51/-0.48 \\
$i$ & 0.96/0.87 & -0.61/-0.22 & 0.99/0.97 & -0.79/-0.11 & 0.81/0.91 & -0.26/0.18 & 0.71/0.97 & -0.42/-0.47 \\
$z$ & \ \ *\ \ /0.89 & \ \ \ *\ \ /-0.11 & \ \ *\ \ /0.95 & \ \ \ *\ \ /\ 0.23 & 0.82/0.92 & -0.31/0.13 & 0.64/0.96 & -0.41/-0.52 \\
$TESS$ & 0.87/\ \ *\ \ \ \  & -0.54/\ \ *\ \ \ \ \  & \ \ *\ \ /\ \ *\ \ \ \  & \ \ *\ \ /\ \ *\ \ \ \  & 0.66/0.75 & 0.17/-0.15 & 0.69/0.86 & -0.55/-0.43
\enddata
\tablecomments{Each entry has two values, one for E1, the other for E2. Entries with $^*$ indicate an insufficient number of observations ($<$100) to determine $Q$ or $M$ for that particular light curve.}
\end{deluxetable*}

In general, our light curves tend to cluster towards the aperiodic-burster region of $QM$ space, though many are symmetric and some are quasi-periodic. In Epoch 1, TW~Hya is aperiodic at 3.54 days ($Q$ for all filters ranges between $[0.87, 0.97]$), but is firmly in the ``burster" regime ($M$ for all filters ranged between $[-0.85, -0.54]$). In Epoch 2, the light curves become more symmetric ($M\in[-0.22, -0.06]$) and straddle the line between quasi- and aperiodic ($Q\in[0.85, 0.89]$). RU~Lup is firmly classified as aperiodic in all its light curves ($Q>$0.95). In E1, it is classified as a burster ($M\in[-0.79, -0.58]$) but in E2 it is more symmetric, even ``dipper-like" in $u$ ($M\in[-0.22, 0.25]$). With the exception of $iz/TESS$, BP~Tau is classified as an aperiodic-burster in E1, with $iz$ near the boundary. $TESS$ is however firmly in the ``quasi-periodic-symmetric" regime, with $Q$=0.66 and M=0.17. The light curves from E2 are equally aperiodic ($Q>$0.75) and show large spread in $M$, ($[-0.56, 0.18]$), with no clear consensus on BP~Tau's classification. GM~Aur is firmly burst-like in every light curve (except $V$ in both epochs), with $M<$-0.41 in all cases. In E1, all light curves are quasi-periodic, with $Q$ between $[0.59, 0.71]$. In E2 though (ignoring $V$), it is far less periodic, with $Q\in[0.86, 0.98]$, as was shown in our periodogram analysis.

It is important to note that the $Q$ \& $M$ metrics were originally designed for high cadence, regular, high precision light curves such as those from $TESS$ and $K2$. That is not the case here, where all of our light curves (besides $TESS$) are irregularly sampled with an average cadence of 12 hours \textit{at best}. \citet{Hillenbrand2022} show that in such irregularly sampled light curves $Q$ can be artificially reduced, especially when $Q$ is high. 

\section{Discussion} \label{sec: Discussion}

Our light curves reveal various degrees and characteristics of variability in our sample, from fairly ordered, highly variable, and periodic, to largely aperiodic, stochastic, and non-variable. Below we discuss the variability characteristics of each target and compare with previous studies.  We then discuss the lack of time lags in the light curves and finally compare the photometric variability to accretion variability derived in Paper I. 

\subsection{Comparing individual objects with previous work}

For TW~Hya, in E2, we recover a robust period near 3.54 days and see evidence for this period in E1, consistent with its 3.57-day rotational period \citep{Huelamo2008, SiciliaAguilar2023}. This signal is consistent throughout both epochs, and may additionally be stable over a timescale of at least one year. This would be in line with the findings of \citet{Donati2011}, who suggest that a cool spot responsible for a 3.57-day period in TW~Hya can persist for at least 3 years. We also detect a signal near 4.27 days in E1, which is stronger (but broader) than the 3.54-day signal, and 4.13 days in E2, which is notably weaker than the 3.54-day signal. These are similar to what was found in \citet{Siwak2014}, though they did not detect an accompanying period near 3.5 days. Furthermore, we see many weak, short-period signals between 1.0-2.5 days in both epochs. These may simply be noise or may be short-lived, transient signals resulting from intrinsic variability in TW~Hya, which would reinforce that TW~Hya experiences some accretion via short-lived, sporadic, low-latitude tongues. Our findings are largely consistent with \citet{SiciliaAguilar2023}, who find that while the emission line footprints (which are due to accretion and are responsible for the 3.57-day radial velocity periodicity) are very stable in TW~Hya, there is considerable photometric variability due to changes in the size, shape, and temperature of the hotspot.

The most notable characteristic in the light curves of RU~Lup is the significant drop in brightness from E1 to E2, dimmer by about 0.71, 0.46, 0.48, 0.46, and 0.46 magnitudes in $BgVri$, respectively. This decrease in brightness is accompanied by a decrease in the median mass accretion rate, from 16.27\accrateunits to 7.36\accrateunits (see Paper I). A similar 0.5-magnitude decrease in brightness was observed in RU~Lup near MJD$\sim$44100 (August 1978), suggesting RU~Lup may undergo large, repetitive changes in its global accretion structure. Unlike the other targets, RU~Lup sees comparable variability amplitude in all bands. This is consistent with our accretion shock modeling in Paper I, where we find that RU~Lup is dominated by a medium density accretion column, whose spectrum is roughly monochromatic at optical wavelengths.

TW~Hya, BP~Tau, and GM~Aur exhibit much stronger variability in $uBg$ (average RMSE$\sim$0.24) than in $riz$ (RMSE$\sim$0.08). In RU~Lup, the average RMSE dispersion in $uBg$ is about 0.29, comparable to the other targets, while for $riz$ it is about 0.18, more than twice that of the other targets. We suggest this is most likely the result of the dominance of the low and medium density accretion columns (see Paper I), which peak at redder wavelengths, while the higher density columns (which peak in the blue-UV) are generally insignificant in RU~Lup. This would lead to comparatively higher variability in $riz$ compared to the other targets. Additionally, RU~Lup possesses a strong accretion-driven outflow which may variably extinguish more blue light than the assumed A$_V$=0.07 would imply.

We recover the previously established 8.19-day period (here 8.14-day) in BP~Tau in $Vri/TESS$ in E1, but in E2 we recover a similar period (8.31 days) in only $uBg$. In neither epoch do we recover the 7.6-day period seen in other studies of BP~Tau \citep{Vrba1986, Simon1990, Osterloh1996}. We note in Paper I that the accretion parameters are similar between E1 and E2 in BP~Tau: all columns contribute to the accretion and the median accretion rate is similar between E1 and E2.

In E1, GM~Aur is highly variable, presumably driven by its 6-day stellar rotation period and the accretion hotspot moving into/out of view. Similar variability and periodicity has been observed in GM~Aur before \citep{Nature, Bouvier2023}, so this is unsurprising and $u$ photometry from E1 shows strong correlations to the accretion-tracing Pa$\beta$ line \citep{Bouvier2023}. In E2, however, the well-established 6-day period in GM~Aur is notably diminished and is detected with significance in only $ug$. Some weak signal near 6 days can be seen in $Brz$, but it is at or below the 0.1\% false alarm probability and thus should not be considered significant. Additionally, the $TESS$ lightcurve, with high cadence and precision, shows no evidence of periodicity near 6 days. 

GM~Aur does still exhibit intrinsic, non-periodic variability, but it is diminished from E1. In E1, the RMSE $u$ and $i$ magnitudes are 0.36 and 0.08, while in E2 they are 0.27 and 0.04. Such diminished variability could occur if GM~Aur was, at all wavelengths, notably brighter in E2, effectively washing out the accretion variability. But all bands show GM~Aur equally as bright in E2 to within about 0.1 magnitudes, plus our shock modeling indicates that the accretion rate in GM~Aur was largely constant in E2, exhibiting much less variability than E1. 

It may be the case that the accretion flow in E2 was stable and symmetric, producing a largely axisymmetric ring on the stellar surface. Such a scenario could produce the diminished variability and the comparable accretion rates as E1, but would wash out much of the rotational modulation. \cite{Kulkarni2013} show that for stars with low-to-moderate ($<$20\textdegree) magnetic obliquity ($\Theta$, the angle between the magnetic and rotation axes), accretion hotspots can manifest as ring-like shapes, becoming more azimuthally symmetric at lower $\Theta$. The magnetic obliquity in GM~Aur is largely unconstrained and may be very small \citep[$\Theta$=13\textdegree$\pm^{+16}_{-13}$, ][]{McGinnis2020}. Furthermore, \citet{Romanova2012, Romanova2021} show that for any inclination, the shape of the accretion footprint is not necessarily stable over time. This all suggests that the accretion hotspot in GM~Aur undergoes notable variability that affects its multi-wavelength light curves, ranging from strong rotational modulation with time lags in 2019 \citep{Nature}, to strong rotational modulation with no time lags in E1, to weak rotational modulation in E2.

\subsection{Variable Dust Extinction} \label{sec: Variable Dust Extinction}

Each light curve for each target exhibits a ``bluer when brighter" color slope, which could be the result of variable extinction, as opposed to variable accretion. Here we discuss this possibility, in this section assuming that all variability originates from dust extinction. To do so, we first assume that any intervening material is primarily composed of graphite and silicate dust, with abundances of 0.004 and 0.0034 \citep{DAlessio2001}. Next, we use the opacity relationships from \citet{DAlessio2001} to determine a total extinction coefficient, A$_{\lambda}$ as:

\begin{equation}
    \kappa_{Total,\ \lambda} = Z_{S}\kappa_{S,\ \lambda} + Z_{G}\kappa_{G,\ \lambda}
\end{equation}
\begin{equation}
    A_{\lambda} = A_{V} \frac{\kappa_{Total,\ \lambda}}{\kappa_{Total,\ V}}
\end{equation}

where $Z_S$ and $Z_G$ are the abundances of silicates and graphite, respectively, $\kappa_{S\ \lambda}$ and $\kappa_{G\ \lambda}$ are the opacities of silicates and graphite at some wavelength $\lambda$, $\kappa_{Total,\ \lambda}$ is the total opacity at $\lambda$, and $V$ represents V-band, or 5500 {\AA}. 

\begin{figure}[ht]
    \centering
    \includegraphics[width=0.47\textwidth]{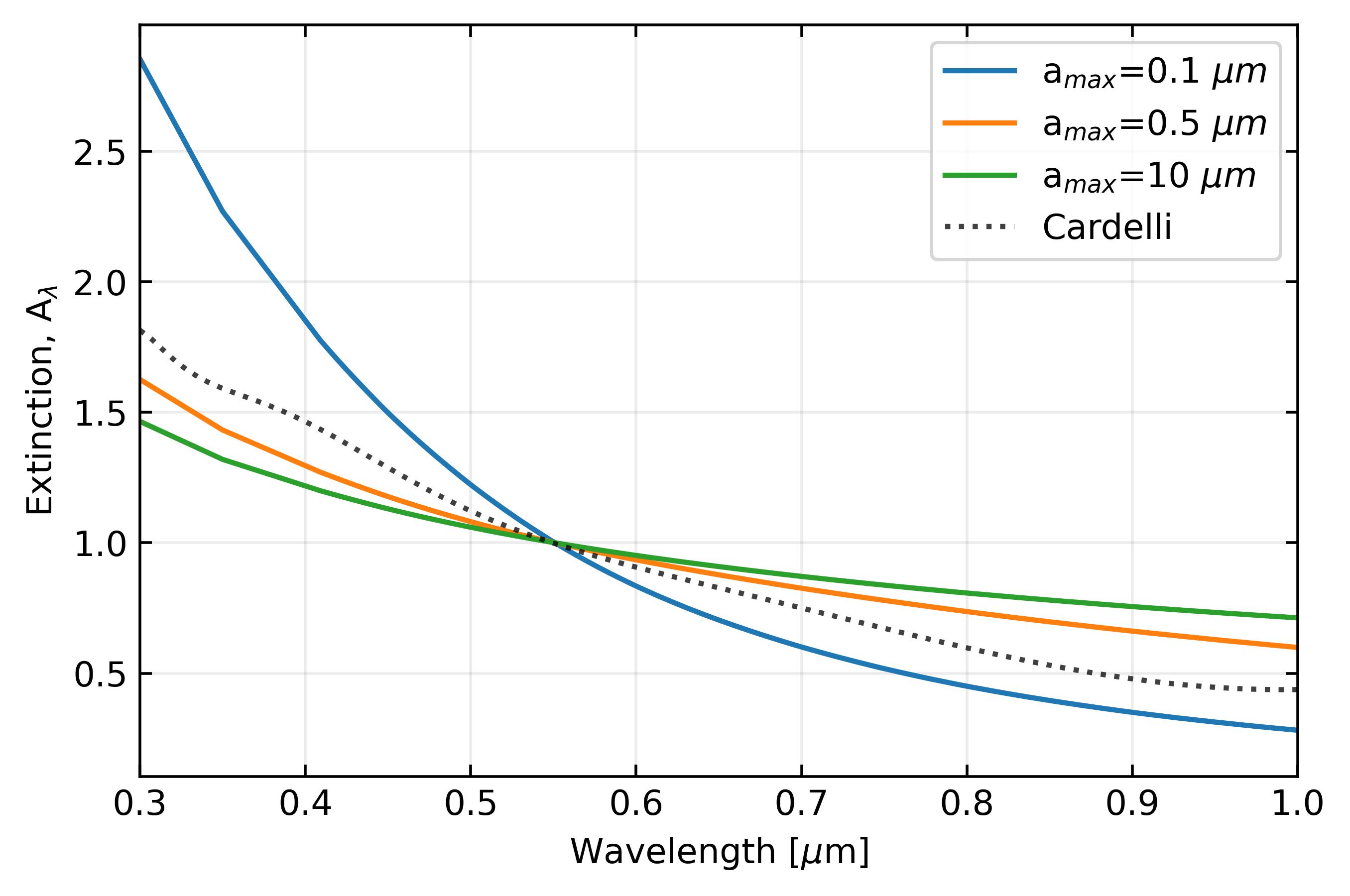}
    \caption{Extinction curves for dust species of various $a_{max}$ (solid lines). Black dotted line is the interstellar extinction law of \citet{Cardelli1989}.}
    \label{fig: a_max Extinction Curves}
\end{figure}

The above opacities depend on the assumed maximum grain size, \amax. We test \amax of 0.1, 0.5, and 10 $\mu$m, assuming a power law distribution of grain sizes with p$^{-3.5}$. The resulting extinction curves for each \amax are shown in Figure \ref{fig: a_max Extinction Curves}, along with the extinction law from \citet{Cardelli1989} which is used in Figure \ref{fig: Light Curve Colors}. For larger grains, the curves are flatter with more monochromatic extinction, while smaller grains preferentially extinct short wavelengths. 

\begin{figure*}
    \centering
    \includegraphics[width=0.975\textwidth]{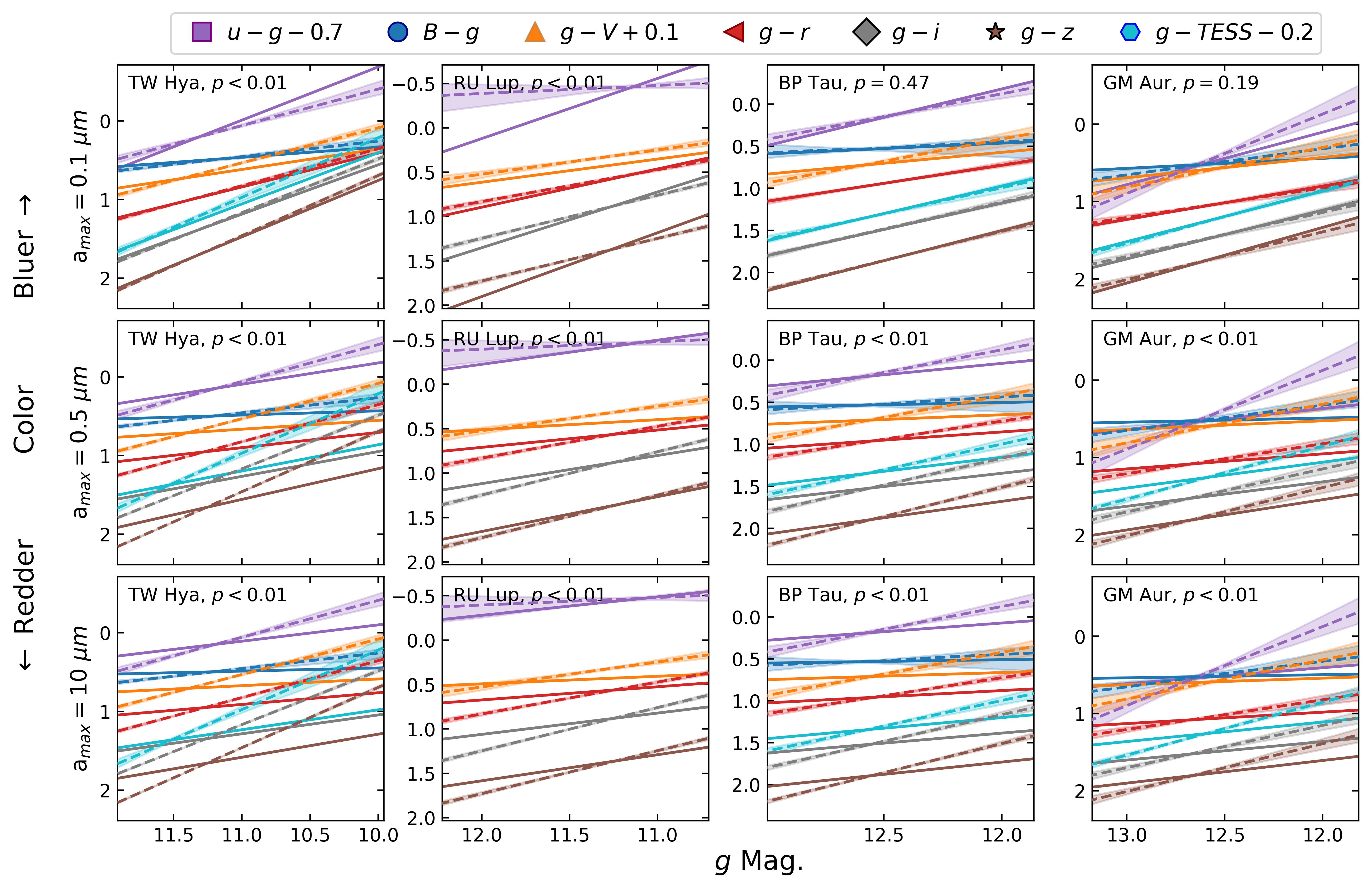}
    \caption{Comparison of observed photometric colors to those expected from local dust extinction. Dashed lines are fitted slopes of photometric colors (dashed lines) with shaded regions denoting uncertainties. Solid lines are slopes expected from a population silicate and graphite dust with size distribution set by \amax. Top to bottom rows are \amax=0.1, 0.5, and 10 $\mu$m. Left to right are TW~Hya, RU~Lup, BP~Tau, GM~Aur. The $p$ value for the given \amax and target are shown in each plot.}
    \label{fig: Extinction Slopes}
\end{figure*}

We use these extinction curves to calculate the expected slope of each photometric color (see Figure \ref{fig: Light Curve Colors}), assuming all color variability originate from dust extinction. Comparing these slopes to those measured from our photometry, we can determine a total $p$ value for the given \amax following \citet{Stouffer1949}. A comparison of the photometric and theoretical slopes are shown in Figure \ref{fig: Extinction Slopes}. In most cases, $p\le0.01$, which indicates that the slope of our photometric colors are not consistent with a dust population for the associated \amax. However, the photometric colors of BP~Tau ($p$=0.47) and GM~Aur($p$=0.19) are not inconsistent with variable extinction from a dust population with \amax=0.1 $\mu$m. 

However, the above assumes that all variability, including bursts above intrinsic brightness, are due to variable dust extinction, which may not be realistic: obscuration from dust should preferentially produce dimming events. To that end, we perform the same analysis on burst-removed light curves. To do so, we first subtract a normalized linear fit from each light curve, then remove any points brighter than the 40th percentile. The same analysis as above reveals that variable dust extinction alone cannot explain the color variability of dimming in TW~Hya or RU~Lup's light curves. Again, though, the color slopes in BP~Tau and GM~Aur are not inconsistent with variable dust extinction. In BP~Tau, we see $p$=0.08 for \amax=0.1 $\mu$m, and in GM~Aur we see $p$=0.18, 0.11, and 0.05 for \amax=0.1, 0.5, and 10 $\mu$m.

This may suggest that a population of small grains contribute to the color variability in these moderately-inclined systems (BP~Tau, GM~Aur). Such a population of small grains may be suggestive of remnant halos or a disk wind \citep{Verhoeff2011, Krijt2011, Olofsson2022}, though a deeper investigation is outside the scope of this work.

\subsection{Time lags in the light curves}

In a 2019 study of GM~Aur, \citet{Nature} reported a time lag in the light curves, where the bluer filters ($ug$) peaked about a day before redder bands ($ri/TESS$). Additionally, they noted that $u$-band emission at one point diminished entirely while the redder bands did not. They interpret this as evidence for a single, large, azimuthally elongated but asymmetric hot spot, where the hotter/denser regions appear earlier in phase as the star rotates. \citet{Robinson2022} also find tentative evidence for smaller, but more general time lags in a sample of 14 CTTSs, where the redder bands lag behind the bluer ones. \citet{Wang2023} observe a phase shift between long/short bandpasses, where $U$-band emission peaks prior to $I$-band emission. They attribute this to the presence of a hot spot that rotates into view prior to a cool spot. Thus, these time lags may be common among CTTSs and would complicate the utility of photometry as a predictor of accretion.

Using the Discrete Correlation Function (DCF) \citep{Edelson1988, Robertson2015}, we attempt to search for evidence of similar time lags. We do not find such evidence in any target, between any two bands. If we assume that all photometric variability originates from the accretion hotspot(s), in the case of GM~Aur, this is likely because the asymmetric hot spot was unstable. Because we still see strong rotational modulation in GM~Aur (at least in E1), it likely possesses a single, large hotspot at moderate latitudes that rotates into/out of view, similar to 2019. Now, however, that hotspot is not asymmetric, leading to synchronous multi-wavelength signals. This is in agreement with the findings of \citet{Bouvier2023} and is consistent with the high-latitude accretion in GM~Aur measured by \citet{McGinnis2020}. In the case of TW~Hya and RU~Lup, their low inclinations would imply that any single, coherent hotspot should generally remain in view even while the star rotates and so we would not expect to see a time lag. BP~Tau is viewed at a moderate inclination, but shows no evidence of time lags. It is thus likely that its accretion hotspot is either symmetric in shape/distribution in the azimuthal direction, or is located at high latitudes and is almost always in view as the star rotates. A high latitude, ever-present hotspot may reinforce the inconsistent/low-significance periodic variability we see in BP~Tau.

It is possible that other effects may have diluted or masked modulation from an asymmetric hotspot. For example, variability in the intrinsic accretion rate (either stochastic or periodic) may wash out the periodic signal from an asymmetric hotspot. If accretion variability was washing out the time lag, it would necessarily have to be in phase with the rotational modulation for several months in both E1 and E2. It is thus unlikely to be the case in BP~Tau or GM~Aur, which show at least moderate rotational periodicity in both epochs.

\subsection{Comments on Q \& M Variability Metrics}

Most of our light curves would be categorized into one of three regimes: quasi-periodic burster (QPB), aperiodic burster (APB), or aperiodic symmetric (APS), with only a few falling into the quasi-periodic symmetric (QPS) regime. All 4 targets are either spread across several regimes (with disagreement amongst the different bandpasses) or change regimes from E1 to E2. In both E1 and E2, BP~Tau's light curves are spread between all 4 of the above regimes. TW~Hya straddles the quasi-periodic regime in both E1 and E2, but is more symmetric and periodic in E2. RU~Lup is always firmly aperiodic, but like TW~Hya is more symmetric in E2. GM~Aur switches from QPB in E1 to APB in E2.

The symmetry metric $M$ is generally lower for bluer filters, with average $M$=-0.41 for $uBg$ versus -0.29 for $riz$, pointing to more burst-like behavior at those shorter wavelengths. This makes sense given that accretion drives most of the variability at these shorter wavelengths, which is always a positive deviation from the mean. $M$ is also lowest in TW~Hya and GM~Aur, which show the strongest accretion variability. There does not appear to be any significant difference in $Q$ between the bluer (median $Q$=0.89) and redder (median $Q$=0.92) bandpasses across all targets. 

We recalculate $Q$ for some alternative periods, either detected in our light curves or from other studies: TW~Hya: 4.28, 6.1 days; RU~Lup: 4.40 days; BP~Tau: 6.1, 7.6 days; GM~Aur: 4.26 days. In most cases these resulted in higher $Q$ values, indicative of less periodic variability on these alternate timescales. The 6.1- and 7.6-day periods for BP~Tau yielded $Q$=0.80 and 0.72 for $TESS$ E2, both considered quasi-periodic, perhaps pointing to some physical mechanism that occurs on these timescales.

\citet{Lin2023} also study the 2021 $TESS$ light curve of BP~Tau, primarily searching for flares, but they characterize it in $QM$ space. They find $Q$=0.83 and $M$=-0.03, in line with \citet{Robinson2022}. We find slightly discrepant values, $Q$=0.66 and $M$=0.17. These differences likely come from the combination of our slightly different processes for calculating $Q/M$ and from our different $TESS$ light curves: we obtained ours using the Python package $TESS$-$Gaia$ Light Curve \citep[$tglc$][]{Han2023} while they utilized the data straight from the SPOC pipeline \citep{Jenkins2016}. They assume a period of 7.6 days instead of our 8.15 days. Regardless, our results suggest that BP~Tau is largely quasi-periodic symmetric, at least as measured by $TESS$. \citet{Robinson2022} measure $Q$ and $M$ for the 2019 $TESS$ light curve of GM~Aur and find $Q$=0.46 and $M$=0.02, showing that GM~Aur has shifted between three different $QM$ regimes (QPS, QPB, APB) from 2019--2022. This may suggest that different phenomena with a comparable importance are responsible for the observed continuum variability in the optical/NIR domain for BP~Tau and GM~Aur.

Overall, our results suggest that the $Q$ and $M$ metrics are dependant on the bandpass used and on time. Bluer bandpasses, which are more sensitive to variable accretion, will likely give lower values of the symmetry metric $M$. Additionally, both $Q$ and $M$ are subject to variability on timescales of a year or greater, where individual targets can switch between variability classes. \citet{Lin2023} also find that variability class can change with time, though on a timescale of 1.6--4 years. Future studies of CTTSs using the $Q$ and $M$ metrics must consider the filter used to obtain the light curves and the potential for variability in individual targets' $QM$ regime.

\subsection{Using photometric variability to infer accretion variability}

The ability to use photometry as a reliable tracer of accretion variability would be highly desirable since it is easier to obtain than spectra, does not require large telescopes, can generally be obtained with high SNR, cadence, and time baselines, and with the ever-expanding collection of meter-class telescopes across the globe, most targets can be observed at nearly any time. However, there are several complications. \citetalias{RE19} show strong correlation between excess $U$-band flux and accretion luminosity in a sample of 5 targets and 25 observations. Their relationships show little scatter and no outliers, suggesting that excess $U$-band flux is a reliable tracer of accretion. Unfortunately, in order to reliably estimate the underlying photosphere, one must know a priori the surface coverage fraction of the accretion hotspot(s). This is difficult without simultaneous UV-optical spectroscopy, which is more difficult to obtain and itself can be used to estimate accretion rates. \citet{Robinson2022} study the relationship between \mdot and $UBVRI/TESS$ photometry. They find that while there is a clear connection between photometry and \mdot, there exists no global relationship between the two. We arrive at a similar conclusion here, as discussed further below.

The plots of \lacc and \mdot over time presented in Figure \ref{fig: Lacc Curves}, while an imperfect representation of the true accretion luminosity/rate of these targets, can help us gauge the extent of the accretion variability. For example, we can see that the accretion in RU~Lup varied more than the limited sampling of $HST$ monitoring would suggest. It peaked at about 34.1\accrateunits in E1 and reached its minimum near $HST$ visit 2.6 at 2.8\accrateunits, a factor of 12.2 (1.09 dex) difference peak-to-peak. TW~Hya reaches very low accretion rates ($\sim$0.04 \accrateunits), though these particularly low values may be due to a combination of our $HST$ data not covering such low accretion rates and some anomalously low photometry points. Regardless, Figure \ref{fig: Lacc Curves} demonstrates that our sample underwent far more accretion variability than our $HST$ monitoring would suggest.

To gauge how well the \lacc and \mdot measured from the light curves reflects the results of our shock modeling, we calculate the Normalized Root Mean Square Error (NRMSE) between the accretion luminosities predicted by our shock model and those predicted by our light curves. To do so, we interpolate the \lacc and \mdot measured from the light curves for the times of the $HST$ observations and calculate the NRMSE between the results of our shock model and those predicted by our light curves for each $HST$ visit. For TW~Hya, RU~Lup, BP~Tau, and GM~Aur, we find NRMSEs of 0.21/0.13, 0.08/0.09, 0.17/0.11, and 0.21/0.12 for E1/E2, respectively. Ultimately, photometry appears to generally be a good predictor of accretion in our sample, but there are caveats. For one, this is not always the case, as seen in BP~Tau, where the relationship breaks down in E1 which may arise from some dust extinction (see Section \ref{sec: Variable Dust Extinction}). Additionally, multi-wavelength light curves are likely necessary for such a technique in order to more accurately reflect the inherently multi-wavelength nature of the accretion flow and hotspot.

As a test of the utility of the total linear relationship between \lacc and \lphot from Figure \ref{fig: Lacc vs Photometry} (solid black line), we attempted to recreate the \lacc and \mdot measured from the light curves using that global relationship. We find NRMSEs of 0.42/0.15, 0.10/0.17, 0.21/0.13, and 0.27/0.15 in E1/E2 for TW~Hya, RU~Lup, BP~Tau, and GM~Aur. Each one is higher than that obtained using intra-object relationships, showing that target-specific relationships between \lacc and \lphotexcess are important for accurate estimates of \lacc from photometry.

\section{Summary} \label{sec: Conclusion}

We conducted a multi-epoch, multi-band photometric monitoring campaign of four CTTSs: TW~Hya, RU~Lup, BP~Tau, and GM~Aur. We analyzed the general light curve characteristics, searched for evidence of periodic modulation, and determined the origin of the variability using photometric colors. Our main findings are as follows:

\begin{enumerate}

    \item All four targets in our sample exhibit strong variability, up to or beyond 1 magnitude in the shortest bands, $u$, $B$, and $g$. Redder bands show less variability, up to about 0.5 magnitudes in $r$, $i$, and $z$.

    \item We generally recover previously established rotational periods in each target except RU~Lup: 3.54 days in TW~Hya, 8.15/8.31 days in BP~Tau, and 6.01 days in GM~Aur. Much of the observed variability is due to rotational modulation. We recover a $\sim$3.54 day period in TW~Hya in both E1 and E2, pointing towards previously observed rotational modulation. We tentatively detect a $\sim$8.15 day period in BP~Tau in both epochs, which we attribute to rotational modulation. In E1 it is recovered by only redder bands ($riz$) while in E2 it is recovered by only bluer bands ($ug$) and slightly longer at 8.31 days. We suggest that a cool spot was present on its surface in E1, while the accretion hotspot dominated in E2. The well-established 6-day rotational period in GM~Aur is recovered in E1, but not in E2, where its overall variability is highly diminished. This change in variability is not accompanied by any notable change in its accretion characteristics, light curve, or line profiles, which may suggest the presence of a highly symmetric ring of accretion in E2. RU~Lup is largely aperiodic in both E1 and E2. 

    \item The light curves of RU~Lup exhibit a significant drop in brightness from E1 to E2, which is accompanied by a decrease in the median mass accretion rate, from 16.27\accrateunits to 7.36\accrateunits, suggesting RU~Lup may undergo large, repetitive changes in its global accretion structure.

    \item Color variability in TW~Hya and RU~Lup is fully inconsistent with either interstellar or local variable extinction, reinforcing variable accretion. In BP~Tau and GM~Aur, the color variability is consistent with variable extinction from small 0.1 $\mu$m grains, suggesting that these small grains may play a role in their variability.

    \item We see strong inter-object correlations between \lacc and excess luminosity above the photosphere that are similar to previous studies, at least in $u^{\prime}$/$U$ bands. These relationships vary per-target and per-bandpass, showing that individual targets exhibit variability trends that differ from the larger CTTSs population. They additionally suggest that our targets (and by extension CTTSs in general) undergo far more accretion variability than a weeks-long monitoring campaign is likely to recover.

    \item Using the $Q$ \& $M$ variability metrics, we find that nearly all of our light curves are considered either quasi-periodic burster, aperiodic burster, or aperiodic symmetric. We find that these classifications can vary on timescales of a year or greater. Additionally, the symmetry metric $M$ is lower (i.e., ``burstier") for bluer bandpasses, reflective of the intrinsic burst-like nature of accretion.

\end{enumerate}

Our study reinforces that CTTSs are highly variable on all timescales, from hours to years, and that the types of variability an individual star exhibits can vary on timescales of about a year. We also show that photometric monitoring can be a useful tool to complement the findings of contemporaneous spectral studies, though the combination of multi-wavelength photometry, moderate cadence ($<$1 day), and months-long baselines are very important to fully understand the accretion variability in an individual CTTS.

\begin{acknowledgments}
This material is based upon work supported by the National Science Foundation under Grant Number AST-2108446. This work is supported by \hst AR-16129 from the Space Telescope Science Institute, which is operated by AURA, Inc. This work benefited from discussions with the ODYSSEUS team (\url{https://sites.bu.edu/odysseus/}); see \cite{Espaillat2022} for an overview of the ODYSSEUS survey. We acknowledge with thanks the variable star observations from the AAVSO International Database contributed by observers worldwide and used in this research. ZG is supported by the ANID FONDECYT Postdoctoral program No. 3220029. This work was also supported by the NKFIH excellence grant TKP2021-NKTA-64.

The operation of the RC80 telescope at Konkoly Observatory has been supported by the GINOP 2.3.2-15-2016-00033 grant of the National Research, Development and Innovation Office (NKFIH) funded by the European Union. We thank Zs\'ofia Bora, Borb\'ala Cseh\', Agoston Horti-D\'avid, Levente Kriskovics, Andr\'as P\'al, \'Ad\'am S\'odor, Zs\'ofia Marianna Szab\'o, R\'obert Szak\'ats, Kriszti\'an Vida, J\'ozsef Vink\'o for their work in these observations.

\end{acknowledgments}

\vspace{5mm}
\facilities{HST, LCOGT, TESS, AAVSO, ZTF, ASAS-SN, Konkoly Observatory, CrAO}

\software{Python, Specutils, AstroPy, pyDCF}

\bibliography{biblio}

\begin{thebibliography}{}
\expandafter\ifx\csname natexlab\endcsname\relax\def\natexlab#1{#1}\fi
\providecommand{\url}[1]{\href{#1}{#1}}
\providecommand{\dodoi}[1]{doi:~\href{http://doi.org/#1}{\nolinkurl{#1}}}
\providecommand{\doeprint}[1]{\href{http://ascl.net/#1}{\nolinkurl{http://ascl.net/#1}}}
\providecommand{\doarXiv}[1]{\href{https://arxiv.org/abs/#1}{\nolinkurl{https://arxiv.org/abs/#1}}}

\bibitem[{{Alcal{\'a}} {et~al.}(2017){Alcal{\'a}}, {Manara}, {Natta}, {Frasca},
  {Testi}, {Nisini}, {Stelzer}, {Williams}, {Antoniucci}, {Biazzo}, {Covino},
  {Esposito}, {Getman}, \& {Rigliaco}}]{Alcala2017}
{Alcal{\'a}}, J.~M., {Manara}, C.~F., {Natta}, A., {et~al.} 2017, \aap, 600,
  A20, \dodoi{10.1051/0004-6361/201629929}

\bibitem[{{Alencar} {et~al.}(2010){Alencar}, {Teixeira}, {Guimar{\~a}es},
  {McGinnis}, {Gameiro}, {Bouvier}, {Aigrain}, {Flaccomio}, \&
  {Favata}}]{Alencar2010}
{Alencar}, S.~H.~P., {Teixeira}, P.~S., {Guimar{\~a}es}, M.~M., {et~al.} 2010,
  \aap, 519, A88, \dodoi{10.1051/0004-6361/201014184}

\bibitem[{{Bellm} {et~al.}(2019){Bellm}, {Kulkarni}, {Graham}, {Dekany},
  {Smith}, {Riddle}, {Masci}, {Helou}, {Prince}, {Adams}, {Barbarino},
  {Barlow}, {Bauer}, {Beck}, {Belicki}, {Biswas}, {Blagorodnova}, {Bodewits},
  {Bolin}, {Brinnel}, {Brooke}, {Bue}, {Bulla}, {Burruss}, {Cenko}, {Chang},
  {Connolly}, {Coughlin}, {Cromer}, {Cunningham}, {De}, {Delacroix}, {Desai},
  {Duev}, {Eadie}, {Farnham}, {Feeney}, {Feindt}, {Flynn}, {Franckowiak},
  {Frederick}, {Fremling}, {Gal-Yam}, {Gezari}, {Giomi}, {Goldstein},
  {Golkhou}, {Goobar}, {Groom}, {Hacopians}, {Hale}, {Henning}, {Ho}, {Hover},
  {Howell}, {Hung}, {Huppenkothen}, {Imel}, {Ip}, {Ivezi{\'c}}, {Jackson},
  {Jones}, {Juric}, {Kasliwal}, {Kaspi}, {Kaye}, {Kelley}, {Kowalski},
  {Kramer}, {Kupfer}, {Landry}, {Laher}, {Lee}, {Lin}, {Lin}, {Lunnan},
  {Giomi}, {Mahabal}, {Mao}, {Miller}, {Monkewitz}, {Murphy}, {Ngeow},
  {Nordin}, {Nugent}, {Ofek}, {Patterson}, {Penprase}, {Porter}, {Rauch},
  {Rebbapragada}, {Reiley}, {Rigault}, {Rodriguez}, {van Roestel}, {Rusholme},
  {van Santen}, {Schulze}, {Shupe}, {Singer}, {Soumagnac}, {Stein}, {Surace},
  {Sollerman}, {Szkody}, {Taddia}, {Terek}, {Van Sistine}, {van Velzen},
  {Vestrand}, {Walters}, {Ward}, {Ye}, {Yu}, {Yan}, \& {Zolkower}}]{ZTF1}
{Bellm}, E.~C., {Kulkarni}, S.~R., {Graham}, M.~J., {et~al.} 2019, \pasp, 131,
  018002, \dodoi{10.1088/1538-3873/aaecbe}

\bibitem[{{Blinova} {et~al.}(2016){Blinova}, {Romanova}, \&
  {Lovelace}}]{Blinova2016}
{Blinova}, A.~A., {Romanova}, M.~M., \& {Lovelace}, R.~V.~E. 2016, \mnras, 459,
  2354, \dodoi{10.1093/mnras/stw786}

\bibitem[{{Bouvier} {et~al.}(2007){Bouvier}, {Alencar}, {Harries},
  {Johns-Krull}, \& {Romanova}}]{Bouvier2007}
{Bouvier}, J., {Alencar}, S.~H.~P., {Harries}, T.~J., {Johns-Krull}, C.~M., \&
  {Romanova}, M.~M. 2007, in Protostars and Planets V, ed. B.~{Reipurth},
  D.~{Jewitt}, \& K.~{Keil}, 479, \dodoi{10.48550/arXiv.astro-ph/0603498}

\bibitem[{{Bouvier} {et~al.}(2023){Bouvier}, {Sousa}, {Pouilly}, {Almenara},
  {Donati}, {Alencar}, {Frasca}, {Grankin}, {Carmona}, {Pantolmos}, {Zaire},
  {Bonfils}, {Bayo}, {Rebull}, {Alonso-Santiago}, {Gameiro}, {Cook}, \&
  {Artigau}}]{Bouvier2023}
{Bouvier}, J., {Sousa}, A., {Pouilly}, K., {et~al.} 2023, \aap, 672, A5,
  \dodoi{10.1051/0004-6361/202245342}

\bibitem[{{Calvet} \& {Gullbring}(1998)}]{Calvet1998}
{Calvet}, N., \& {Gullbring}, E. 1998, \apj, 509, 802, \dodoi{10.1086/306527}

\bibitem[{{Cardelli} {et~al.}(1989){Cardelli}, {Clayton}, \&
  {Mathis}}]{Cardelli1989}
{Cardelli}, J.~A., {Clayton}, G.~C., \& {Mathis}, J.~S. 1989, \apj, 345, 245,
  \dodoi{10.1086/167900}

\bibitem[{{Cody} \& {Hillenbrand}(2018)}]{Cody2018}
{Cody}, A.~M., \& {Hillenbrand}, L.~A. 2018, \aj, 156, 71,
  \dodoi{10.3847/1538-3881/aacead}

\bibitem[{{Cody} {et~al.}(2014){Cody}, {Stauffer}, {Baglin}, {Micela},
  {Rebull}, {Flaccomio}, {Morales-Calder{\'o}n}, {Aigrain}, {Bouvier},
  {Hillenbrand}, {Gutermuth}, {Song}, {Turner}, {Alencar}, {Zwintz},
  {Plavchan}, {Carpenter}, {Findeisen}, {Carey}, {Terebey}, {Hartmann},
  {Calvet}, {Teixeira}, {Vrba}, {Wolk}, {Covey}, {Poppenhaeger}, {G{\"u}nther},
  {Forbrich}, {Whitney}, {Affer}, {Herbst}, {Hora}, {Barrado}, {Holtzman},
  {Marchis}, {Wood}, {Medeiros Guimar{\~a}es}, {Lillo Box}, {Gillen},
  {McQuillan}, {Espaillat}, {Allen}, {D'Alessio}, \& {Favata}}]{Cody2014}
{Cody}, A.~M., {Stauffer}, J., {Baglin}, A., {et~al.} 2014, \aj, 147, 82,
  \dodoi{10.1088/0004-6256/147/4/82}

\bibitem[{{D'Alessio} {et~al.}(2001){D'Alessio}, {Calvet}, \&
  {Hartmann}}]{DAlessio2001}
{D'Alessio}, P., {Calvet}, N., \& {Hartmann}, L. 2001, \apj, 553, 321,
  \dodoi{10.1086/320655}

\bibitem[{{Donati} {et~al.}(2011){Donati}, {Gregory}, {Alencar}, {Bouvier},
  {Hussain}, {Skelly}, {Dougados}, {Jardine}, {M{\'e}nard}, {Romanova}, \&
  {Unruh}}]{Donati2011}
{Donati}, J.~F., {Gregory}, S.~G., {Alencar}, S.~H.~P., {et~al.} 2011, \mnras,
  417, 472, \dodoi{10.1111/j.1365-2966.2011.19288.x}

\bibitem[{{Dupree}(2013)}]{Dupree2013}
{Dupree}, A.~K. 2013, Astronomische Nachrichten, 334, 73,
  \dodoi{10.1002/asna.201211748}

\bibitem[{{Edelson} \& {Krolik}(1988)}]{Edelson1988}
{Edelson}, R.~A., \& {Krolik}, J.~H. 1988, \apj, 333, 646,
  \dodoi{10.1086/166773}

\bibitem[{{Espaillat} {et~al.}(2021){Espaillat}, {Robinson}, {Romanova},
  {Thanathibodee}, {Wendeborn}, {Calvet}, {Reynolds}, \& {Muzerolle}}]{Nature}
{Espaillat}, C.~C., {Robinson}, C.~E., {Romanova}, M.~M., {et~al.} 2021, \nat,
  597, 41, \dodoi{10.1038/s41586-021-03751-5}

\bibitem[{{Espaillat} {et~al.}(2022){Espaillat}, {Herczeg}, {Thanathibodee},
  {Pittman}, {Calvet}, {Arulanantham}, {France}, {Serna}, {Hern{\'a}ndez},
  {K{\'o}sp{\'a}l}, {Walter}, {Frasca}, {Fischer}, {Johns-Krull}, {Schneider},
  {Robinson}, {Edwards}, {{\'A}brah{\'a}m}, {Fang}, {Erkal}, {Manara},
  {Alcal{\'a}}, {Alecian}, {Alexander}, {Alonso-Santiago}, {Antoniucci},
  {Ardila}, {Banzatti}, {Benisty}, {Bergin}, {Biazzo}, {Brice{\~n}o},
  {Campbell-White}, {Cleeves}, {Coffey}, {Eisl{\"o}ffel}, {Facchini}, {Fedele},
  {Fiorellino}, {Froebrich}, {Gangi}, {Giannini}, {Grankin}, {G{\"u}nther},
  {Guo}, {Hartmann}, {Hillenbrand}, {Hinton}, {Kastner}, {Koen}, {Mauc{\'o}},
  {Mendigut{\'\i}a}, {Nisini}, {Panwar}, {Principe}, {Robberto},
  {Sicilia-Aguilar}, {Valenti}, {Wendeborn}, {Williams}, {Xu}, \&
  {Yadav}}]{Espaillat2022}
{Espaillat}, C.~C., {Herczeg}, G.~J., {Thanathibodee}, T., {et~al.} 2022, \aj,
  163, 114, \dodoi{10.3847/1538-3881/ac479d}

\bibitem[{{Fallscheer} \& {Herbst}(2006)}]{Fallscheer2006}
{Fallscheer}, C., \& {Herbst}, W. 2006, \apjl, 647, L155,
  \dodoi{10.1086/507525}

\bibitem[{{Fischer} {et~al.}(2023){Fischer}, {Hillenbrand}, {Herczeg},
  {Johnstone}, {Kospal}, \& {Dunham}}]{PPVII10}
{Fischer}, W.~J., {Hillenbrand}, L.~A., {Herczeg}, G.~J., {et~al.} 2023, in
  Astronomical Society of the Pacific Conference Series, Vol. 534, Protostars
  and Planets VII, ed. S.~{Inutsuka}, Y.~{Aikawa}, T.~{Muto}, K.~{Tomida}, \&
  M.~{Tamura}, 355, \dodoi{10.48550/arXiv.2203.11257}

\bibitem[{{Flaischlen} {et~al.}(2022){Flaischlen}, {Preibisch}, {Kluge},
  {Manara}, \& {Ercolano}}]{Flaischlen2022}
{Flaischlen}, S., {Preibisch}, T., {Kluge}, M., {Manara}, C.~F., \& {Ercolano},
  B. 2022, \aap, 666, A55, \dodoi{10.1051/0004-6361/202142630}

\bibitem[{{Giovannelli} {et~al.}(1995){Giovannelli}, {Vittone}, {Rossi},
  {Errico}, {Bisnovatyi-Kogan}, {Kurt}, {Lamzin}, {Larionov}, {Sheffer}, \&
  {Sidorenkov}}]{Giovannelli1995}
{Giovannelli}, F., {Vittone}, A.~A., {Rossi}, C., {et~al.} 1995, \aaps, 114,
  341

\bibitem[{{Gullbring} {et~al.}(1996){Gullbring}, {Barwig}, {Chen}, {Gahm}, \&
  {Bao}}]{Gullbring1996}
{Gullbring}, E., {Barwig}, H., {Chen}, P.~S., {Gahm}, G.~F., \& {Bao}, M.~X.
  1996, \aap, 307, 791

\bibitem[{{Gullbring} {et~al.}(1998){Gullbring}, {Hartmann}, {Brice{\~n}o}, \&
  {Calvet}}]{Gullbring1998}
{Gullbring}, E., {Hartmann}, L., {Brice{\~n}o}, C., \& {Calvet}, N. 1998, \apj,
  492, 323, \dodoi{10.1086/305032}

\bibitem[{{Guo} {et~al.}(2018){Guo}, {Herczeg}, {Jose}, {Fu}, {Chiang},
  {Grankin}, {Michel}, {Kesh Yadav}, {Liu}, {Chen}, {Li}, {Xue}, {Niu},
  {Subramaniam}, {Sharma}, {Prasert}, {Flores-Fajardo}, {Castro}, \&
  {Altamirano}}]{Guo2018}
{Guo}, Z., {Herczeg}, G.~J., {Jose}, J., {et~al.} 2018, \apj, 852, 56,
  \dodoi{10.3847/1538-4357/aa9e52}

\bibitem[{{Han} \& {Brandt}(2023{\natexlab{a}})}]{tglc}
{Han}, T., \& {Brandt}, T.~D. 2023{\natexlab{a}}, \aj, 165, 71,
  \dodoi{10.3847/1538-3881/acaaa7}

\bibitem[{{Han} \& {Brandt}(2023{\natexlab{b}})}]{Han2023}
---. 2023{\natexlab{b}}, \aj, 165, 71, \dodoi{10.3847/1538-3881/acaaa7}

\bibitem[{{Hartmann} {et~al.}(2016){Hartmann}, {Herczeg}, \&
  {Calvet}}]{Hartmann2016}
{Hartmann}, L., {Herczeg}, G., \& {Calvet}, N. 2016, \araa, 54, 135,
  \dodoi{10.1146/annurev-astro-081915-023347}

\bibitem[{{Henden} {et~al.}(2015){Henden}, {Levine}, {Terrell}, \&
  {Welch}}]{APASS}
{Henden}, A.~A., {Levine}, S., {Terrell}, D., \& {Welch}, D.~L. 2015, in
  American Astronomical Society Meeting Abstracts, Vol. 225, American
  Astronomical Society Meeting Abstracts \#225, 336.16

\bibitem[{{Herczeg} {et~al.}(2023){Herczeg}, {Chen}, {Donati}, {Dupree},
  {Walter}, {Hillenbrand}, {Johns-Krull}, {Manara}, {G{\"u}nther}, {Fang},
  {Schneider}, {Valenti}, {Alencar}, {Venuti}, {Alcal{\'a}}, {Frasca},
  {Arulanantham}, {Linsky}, {Bouvier}, {Brickhouse}, {Calvet}, {Espaillat},
  {Campbell-White}, {Carpenter}, {Chang}, {Cruz}, {Dahm}, {Eisl{\"o}ffel},
  {Edwards}, {Fischer}, {Guo}, {Henning}, {Ji}, {Jose}, {Kastner}, {Launhardt},
  {Principe}, {Robinson}, {Serna}, {Siwak}, {Sterzik}, \&
  {Takasao}}]{Herczeg2023}
{Herczeg}, G.~J., {Chen}, Y., {Donati}, J.-F., {et~al.} 2023, \apj, 956, 102,
  \dodoi{10.3847/1538-4357/acf468}

\bibitem[{{Hillenbrand} {et~al.}(2022){Hillenbrand}, {Kiker}, {Gee}, {Lester},
  {Braunfeld}, {Rebull}, \& {Kuhn}}]{Hillenbrand2022}
{Hillenbrand}, L.~A., {Kiker}, T.~J., {Gee}, M., {et~al.} 2022, \aj, 163, 263,
  \dodoi{10.3847/1538-3881/ac62d8}

\bibitem[{{Hinton} {et~al.}(2022){Hinton}, {France}, {Batista}, {Serna},
  {Hern{\'a}ndez}, {G{\"u}nther}, {Kowalski}, \& {Schneider}}]{Hinton2022}
{Hinton}, P.~C., {France}, K., {Batista}, M.~G., {et~al.} 2022, \apj, 939, 82,
  \dodoi{10.3847/1538-4357/ac8f26}

\bibitem[{{Hu{\'e}lamo} {et~al.}(2008){Hu{\'e}lamo}, {Figueira}, {Bonfils},
  {Santos}, {Pepe}, {Gillon}, {Azevedo}, {Barman}, {Fern{\'a}ndez}, {di Folco},
  {Guenther}, {Lovis}, {Melo}, {Queloz}, \& {Udry}}]{Huelamo2008}
{Hu{\'e}lamo}, N., {Figueira}, P., {Bonfils}, X., {et~al.} 2008, \aap, 489, L9,
  \dodoi{10.1051/0004-6361:200810596}

\bibitem[{{Ingleby} {et~al.}(2013){Ingleby}, {Calvet}, {Herczeg}, {Blaty},
  {Walter}, {Ardila}, {Alexander}, {Edwards}, {Espaillat}, {Gregory},
  {Hillenbrand}, \& {Brown}}]{Ingleby2013}
{Ingleby}, L., {Calvet}, N., {Herczeg}, G., {et~al.} 2013, \apj, 767, 112,
  \dodoi{10.1088/0004-637X/767/2/112}

\bibitem[{{Jenkins} {et~al.}(2016){Jenkins}, {Twicken}, {McCauliff},
  {Campbell}, {Sanderfer}, {Lung}, {Mansouri-Samani}, {Girouard}, {Tenenbaum},
  {Klaus}, {Smith}, {Caldwell}, {Chacon}, {Henze}, {Heiges}, {Latham},
  {Morgan}, {Swade}, {Rinehart}, \& {Vanderspek}}]{Jenkins2016}
{Jenkins}, J.~M., {Twicken}, J.~D., {McCauliff}, S., {et~al.} 2016, in Society
  of Photo-Optical Instrumentation Engineers (SPIE) Conference Series, Vol.
  9913, Software and Cyberinfrastructure for Astronomy IV, ed. G.~{Chiozzi} \&
  J.~C. {Guzman}, 99133E, \dodoi{10.1117/12.2233418}

\bibitem[{{Kochanek} {et~al.}(2017){Kochanek}, {Shappee}, {Stanek}, {Holoien},
  {Thompson}, {Prieto}, {Dong}, {Shields}, {Will}, {Britt}, {Perzanowski}, \&
  {Pojma{\'n}ski}}]{ASAS-SN2}
{Kochanek}, C.~S., {Shappee}, B.~J., {Stanek}, K.~Z., {et~al.} 2017, \pasp,
  129, 104502, \dodoi{10.1088/1538-3873/aa80d9}

\bibitem[{{Krijt} \& {Dominik}(2011)}]{Krijt2011}
{Krijt}, S., \& {Dominik}, C. 2011, \aap, 531, A80,
  \dodoi{10.1051/0004-6361/201116757}

\bibitem[{{Kulkarni} \& {Romanova}(2008)}]{Kulkarni2008}
{Kulkarni}, A.~K., \& {Romanova}, M.~M. 2008, \mnras, 386, 673,
  \dodoi{10.1111/j.1365-2966.2008.13094.x}

\bibitem[{{Kulkarni} \& {Romanova}(2013)}]{Kulkarni2013}
---. 2013, \mnras, 433, 3048, \dodoi{10.1093/mnras/stt945}

\bibitem[{{Kurosawa} \& {Romanova}(2013)}]{Kurosawa2013}
{Kurosawa}, R., \& {Romanova}, M.~M. 2013, \mnras, 431, 2673,
  \dodoi{10.1093/mnras/stt365}

\bibitem[{{Lasker} {et~al.}(2008){Lasker}, {Lattanzi}, {McLean}, {Bucciarelli},
  {Drimmel}, {Garcia}, {Greene}, {Guglielmetti}, {Hanley}, {Hawkins},
  {Laidler}, {Loomis}, {Meakes}, {Mignani}, {Morbidelli}, {Morrison},
  {Pannunzio}, {Rosenberg}, {Sarasso}, {Smart}, {Spagna}, {Sturch},
  {Volpicelli}, {White}, {Wolfe}, \& {Zacchei}}]{GSC}
{Lasker}, B.~M., {Lattanzi}, M.~G., {McLean}, B.~J., {et~al.} 2008, \aj, 136,
  735, \dodoi{10.1088/0004-6256/136/2/735}

\bibitem[{{Lin} {et~al.}(2023){Lin}, {Ip}, {Hsiao}, {Chang}, {Song}, \&
  {Luo}}]{Lin2023}
{Lin}, C.-L., {Ip}, W.-H., {Hsiao}, Y., {et~al.} 2023, \aj, 166, 82,
  \dodoi{10.3847/1538-3881/ace322}

\bibitem[{{Manara} {et~al.}(2013){Manara}, {Testi}, {Rigliaco}, {Alcal{\'a}},
  {Natta}, {Stelzer}, {Biazzo}, {Covino}, {Covino}, {Cupani}, {D'Elia}, \&
  {Randich}}]{Manara2013}
{Manara}, C.~F., {Testi}, L., {Rigliaco}, E., {et~al.} 2013, \aap, 551, A107,
  \dodoi{10.1051/0004-6361/201220921}

\bibitem[{{Manara} {et~al.}(2021){Manara}, {Frasca}, {Venuti}, {Siwak},
  {Herczeg}, {Calvet}, {Hernandez}, {Tychoniec}, {Gangi}, {Alcal{\'a}},
  {Boffin}, {Nisini}, {Robberto}, {Briceno}, {Campbell-White},
  {Sicilia-Aguilar}, {McGinnis}, {Fedele}, {K{\'o}sp{\'a}l}, {{\'A}brah{\'a}m},
  {Alonso-Santiago}, {Antoniucci}, {Arulanantham}, {Bacciotti}, {Banzatti},
  {Beccari}, {Benisty}, {Biazzo}, {Bouvier}, {Cabrit}, {Caratti o Garatti},
  {Coffey}, {Covino}, {Dougados}, {Eisl{\"o}ffel}, {Ercolano}, {Espaillat},
  {Erkal}, {Facchini}, {Fang}, {Fiorellino}, {Fischer}, {France}, {Gameiro},
  {Garcia Lopez}, {Giannini}, {Ginski}, {Grankin}, {G{\"u}nther}, {Hartmann},
  {Hillenbrand}, {Hussain}, {James}, {Koutoulaki}, {Lodato}, {Mauc{\'o}},
  {Mendigut{\'\i}a}, {Mentel}, {Miotello}, {Oudmaijer}, {Rigliaco}, {Rosotti},
  {Sanchis}, {Schneider}, {Spina}, {Stelzer}, {Testi}, {Thanathibodee}, {Vink},
  {Walter}, {Williams}, \& {Zsidi}}]{Manara2021}
{Manara}, C.~F., {Frasca}, A., {Venuti}, L., {et~al.} 2021, \aap, 650, A196,
  \dodoi{10.1051/0004-6361/202140639}

\bibitem[{{McCully} {et~al.}(2018){McCully}, {Volgenau}, {Harbeck}, {Lister},
  {Saunders}, {Turner}, {Siiverd}, \& {Bowman}}]{BANZAI}
{McCully}, C., {Volgenau}, N.~H., {Harbeck}, D.-R., {et~al.} 2018, in Society
  of Photo-Optical Instrumentation Engineers (SPIE) Conference Series, Vol.
  10707, Software and Cyberinfrastructure for Astronomy V, ed. J.~C. {Guzman}
  \& J.~{Ibsen}, 107070K, \dodoi{10.1117/12.2314340}

\bibitem[{{McGinnis} {et~al.}(2020){McGinnis}, {Bouvier}, \&
  {Gallet}}]{McGinnis2020}
{McGinnis}, P., {Bouvier}, J., \& {Gallet}, F. 2020, \mnras, 497, 2142,
  \dodoi{10.1093/mnras/staa2041}

\bibitem[{{Olofsson} {et~al.}(2022){Olofsson}, {Th{\'e}bault}, {Kennedy}, \&
  {Bayo}}]{Olofsson2022}
{Olofsson}, J., {Th{\'e}bault}, P., {Kennedy}, G.~M., \& {Bayo}, A. 2022, \aap,
  664, A122, \dodoi{10.1051/0004-6361/202243794}

\bibitem[{{Osterloh} {et~al.}(1996){Osterloh}, {Thommes}, \&
  {Kania}}]{Osterloh1996}
{Osterloh}, M., {Thommes}, E., \& {Kania}, U. 1996, VizieR Online Data Catalog,
  J/A+AS/120/267

\bibitem[{{Percy} {et~al.}(2010){Percy}, {Esteves}, {Glasheen}, {Lin}, {Long},
  {Mashintsova}, {Terziev}, \& {Wu}}]{Percy2010}
{Percy}, J.~R., {Esteves}, S., {Glasheen}, J., {et~al.} 2010, \jaavso, 38, 151

\bibitem[{{Percy} {et~al.}(2006){Percy}, {Gryc}, {Wong}, \&
  {Herbst}}]{Percy2006b}
{Percy}, J.~R., {Gryc}, W.~K., {Wong}, J. C.~Y., \& {Herbst}, W. 2006, \pasp,
  118, 1390, \dodoi{10.1086/508557}

\bibitem[{{Percy} \& {Palaniappan}(2006)}]{Percy2006a}
{Percy}, J.~R., \& {Palaniappan}, R. 2006, \jaavso, 35, 290

\bibitem[{{Robertson} {et~al.}(2015){Robertson}, {Gallo}, {Zoghbi}, \&
  {Fabian}}]{Robertson2015}
{Robertson}, D.~R.~S., {Gallo}, L.~C., {Zoghbi}, A., \& {Fabian}, A.~C. 2015,
  \mnras, 453, 3455, \dodoi{10.1093/mnras/stv1575}

\bibitem[{{Robinson} \& {Espaillat}(2019)}]{RE19}
{Robinson}, C.~E., \& {Espaillat}, C.~C. 2019, \apj, 874, 129,
  \dodoi{10.3847/1538-4357/ab0d8d}

\bibitem[{{Robinson} {et~al.}(2021){Robinson}, {Espaillat}, \&
  {Owen}}]{Robinson2021}
{Robinson}, C.~E., {Espaillat}, C.~C., \& {Owen}, J.~E. 2021, \apj, 908, 16,
  \dodoi{10.3847/1538-4357/abd410}

\bibitem[{{Robinson} {et~al.}(2022){Robinson}, {Espaillat}, \&
  {Rodriguez}}]{Robinson2022}
{Robinson}, C.~E., {Espaillat}, C.~C., \& {Rodriguez}, J.~E. 2022, \apj, 935,
  54, \dodoi{10.3847/1538-4357/ac7e51}

\bibitem[{{Roman-Duval} {et~al.}(2020){Roman-Duval}, {Proffitt}, {Taylor},
  {Monroe}, {Fischer}, {Fischer}, {Fullerton}, {Aloisi}, {Britt}, {Busko},
  {Carlberg}, {De Rosa}, {Jedrzejewski}, {Lockwood}, {Frazer}, {Hernandez},
  {James}, {Oliveira}, {Plesha}, {Riedel}, {Riley}, {Sahnow}, {Sankrit},
  {Shaw}, {Smith}, {Sohn}, {Som}, {Ubeda}, \& {Welty}}]{ULYSSES}
{Roman-Duval}, J., {Proffitt}, C.~R., {Taylor}, J.~M., {et~al.} 2020, Research
  Notes of the American Astronomical Society, 4, 205,
  \dodoi{10.3847/2515-5172/abca2f}

\bibitem[{{Romanova} {et~al.}(2021){Romanova}, {Koldoba}, {Ustyugova},
  {Blinova}, {Lai}, \& {Lovelace}}]{Romanova2021}
{Romanova}, M.~M., {Koldoba}, A.~V., {Ustyugova}, G.~V., {et~al.} 2021, \mnras,
  506, 372, \dodoi{10.1093/mnras/stab1724}

\bibitem[{{Romanova} {et~al.}(2008){Romanova}, {Kulkarni}, \&
  {Lovelace}}]{Romanova2008}
{Romanova}, M.~M., {Kulkarni}, A.~K., \& {Lovelace}, R. V.~E. 2008, \apjl, 673,
  L171, \dodoi{10.1086/527298}

\bibitem[{{Romanova} {et~al.}(2004){Romanova}, {Ustyugova}, {Koldoba}, \&
  {Lovelace}}]{Romanova2004}
{Romanova}, M.~M., {Ustyugova}, G.~V., {Koldoba}, A.~V., \& {Lovelace},
  R.~V.~E. 2004, \apjl, 616, L151, \dodoi{10.1086/426586}

\bibitem[{{Romanova} {et~al.}(2012){Romanova}, {Ustyugova}, {Koldoba}, \&
  {Lovelace}}]{Romanova2012}
---. 2012, \mnras, 421, 63, \dodoi{10.1111/j.1365-2966.2011.20055.x}

\bibitem[{{Shappee} {et~al.}(2014){Shappee}, {Prieto}, {Grupe}, {Kochanek},
  {Stanek}, {De Rosa}, {Mathur}, {Zu}, {Peterson}, {Pogge}, {Komossa}, {Im},
  {Jencson}, {Holoien}, {Basu}, {Beacom}, {Szczygie{\l}}, {Brimacombe},
  {Adams}, {Campillay}, {Choi}, {Contreras}, {Dietrich}, {Dubberley},
  {Elphick}, {Foale}, {Giustini}, {Gonzalez}, {Hawkins}, {Howell}, {Hsiao},
  {Koss}, {Leighly}, {Morrell}, {Mudd}, {Mullins}, {Nugent}, {Parrent},
  {Phillips}, {Pojmanski}, {Rosing}, {Ross}, {Sand}, {Terndrup}, {Valenti},
  {Walker}, \& {Yoon}}]{ASAS-SN1}
{Shappee}, B.~J., {Prieto}, J.~L., {Grupe}, D., {et~al.} 2014, \apj, 788, 48,
  \dodoi{10.1088/0004-637X/788/1/48}

\bibitem[{{Sicilia-Aguilar} {et~al.}(2023){Sicilia-Aguilar}, {Campbell-White},
  {Roccatagliata}, {Desira}, {Gregory}, {Scholz}, {Fang}, {Cruz-Saenz de
  Miera}, {K{\'o}sp{\'a}l}, {Matsumura}, \&
  {{\'A}brah{\'a}m}}]{SiciliaAguilar2023}
{Sicilia-Aguilar}, A., {Campbell-White}, J., {Roccatagliata}, V., {et~al.}
  2023, \mnras, 526, 4885, \dodoi{10.1093/mnras/stad3029}

\bibitem[{{Simon} {et~al.}(1990){Simon}, {Vrba}, \& {Herbst}}]{Simon1990}
{Simon}, T., {Vrba}, F.~J., \& {Herbst}, W. 1990, \aj, 100, 1957,
  \dodoi{10.1086/115651}

\bibitem[{{Siwak} {et~al.}(2011){Siwak}, {Rucinski}, {Matthews},
  {Pojma{\'n}ski}, {Kuschnig}, {Guenther}, {Moffat}, {Sasselov}, \&
  {Weiss}}]{Siwak2011}
{Siwak}, M., {Rucinski}, S.~M., {Matthews}, J.~M., {et~al.} 2011, \mnras, 410,
  2725, \dodoi{10.1111/j.1365-2966.2010.17649.x}

\bibitem[{{Siwak} {et~al.}(2014){Siwak}, {Rucinski}, {Matthews}, {Guenther},
  {Kuschnig}, {Moffat}, {Rowe}, {Sasselov}, \& {Weiss}}]{Siwak2014}
---. 2014, \mnras, 444, 327, \dodoi{10.1093/mnras/stu1304}

\bibitem[{{Siwak} {et~al.}(2016){Siwak}, {Ogloza}, {Rucinski}, {Moffat},
  {Matthews}, {Cameron}, {Guenther}, {Kuschnig}, {Rowe}, {Sasselov}, \&
  {Weiss}}]{Siwak2016}
{Siwak}, M., {Ogloza}, W., {Rucinski}, S.~M., {et~al.} 2016, \mnras, 456, 3972,
  \dodoi{10.1093/mnras/stv2848}

\bibitem[{{Siwak} {et~al.}(2018){Siwak}, {Ogloza}, {Moffat}, {Matthews},
  {Rucinski}, {Kallinger}, {Kuschnig}, {Cameron}, {Weiss}, {Rowe}, {Guenther},
  \& {Sasselov}}]{Siwak2018}
{Siwak}, M., {Ogloza}, W., {Moffat}, A. F.~J., {et~al.} 2018, \mnras, 478, 758,
  \dodoi{10.1093/mnras/sty1220}

\bibitem[{{Stelzer} {et~al.}(2013){Stelzer}, {Frasca}, {Alcal{\'a}}, {Manara},
  {Biazzo}, {Covino}, {Rigliaco}, {Testi}, {Covino}, \& {D'Elia}}]{Stelzer2013}
{Stelzer}, B., {Frasca}, A., {Alcal{\'a}}, J.~M., {et~al.} 2013, \aap, 558,
  A141, \dodoi{10.1051/0004-6361/201321979}

\bibitem[{{Stempels} {et~al.}(2007){Stempels}, {Gahm}, \&
  {Petrov}}]{Stempels2007}
{Stempels}, H.~C., {Gahm}, G.~F., \& {Petrov}, P.~P. 2007, \aap, 461, 253,
  \dodoi{10.1051/0004-6361:20065268}

\bibitem[{{Stouffer} {et~al.}(1949){Stouffer}, {Suchman}, {Devinney}, {Star},
  \& {Williams}}]{Stouffer1949}
{Stouffer}, S.~A., {Suchman}, E.~A., {Devinney}, L.~C., {Star}, S.~A., \&
  {Williams}, R.~M., J. 1949, {The American soldier: Adjustment during army
  life. (Studies in social psychology in World War II} ({Princeton Univ.
  Press})

\bibitem[{{Tonry} {et~al.}(2018){Tonry}, {Denneau}, {Flewelling}, {Heinze},
  {Onken}, {Smartt}, {Stalder}, {Weiland}, \& {Wolf}}]{ATLAS-REFCAT2}
{Tonry}, J.~L., {Denneau}, L., {Flewelling}, H., {et~al.} 2018, \apj, 867, 105,
  \dodoi{10.3847/1538-4357/aae386}

\bibitem[{{VanderPlas}(2018)}]{VanderPlas2018}
{VanderPlas}, J.~T. 2018, \apjs, 236, 16, \dodoi{10.3847/1538-4365/aab766}

\bibitem[{{Venuti} {et~al.}(2015){Venuti}, {Bouvier}, {Irwin}, {Stauffer},
  {Hillenbrand}, {Rebull}, {Cody}, {Alencar}, {Micela}, {Flaccomio}, \&
  {Peres}}]{Venuti2015}
{Venuti}, L., {Bouvier}, J., {Irwin}, J., {et~al.} 2015, \aap, 581, A66,
  \dodoi{10.1051/0004-6361/201526164}

\bibitem[{{Verhoeff} {et~al.}(2011){Verhoeff}, {Min}, {Pantin}, {Waters},
  {Tielens}, {Honda}, {Fujiwara}, {Bouwman}, {van Boekel}, {Dougherty}, {de
  Koter}, {Dominik}, \& {Mulders}}]{Verhoeff2011}
{Verhoeff}, A.~P., {Min}, M., {Pantin}, E., {et~al.} 2011, \aap, 528, A91,
  \dodoi{10.1051/0004-6361/201014952}

\bibitem[{{Vrba} {et~al.}(1986){Vrba}, {Rydgren}, {Chugainov}, {Shakovskaia},
  \& {Zak}}]{Vrba1986}
{Vrba}, F.~J., {Rydgren}, A.~E., {Chugainov}, P.~F., {Shakovskaia}, N.~I., \&
  {Zak}, D.~S. 1986, \apj, 306, 199, \dodoi{10.1086/164332}

\bibitem[{{Wang} {et~al.}(2023){Wang}, {Herczeg}, {Liu}, {Fang}, {Johnstone},
  {Lee}, {Walter}, {Hambsch}, {Contreras Pe{\~n}a}, {Lee}, {Millward},
  {Pearce}, {Monard}, \& {Zhou}}]{Wang2023}
{Wang}, M.-T., {Herczeg}, G.~J., {Liu}, H.-G., {et~al.} 2023, \apj, 957, 113,
  \dodoi{10.3847/1538-4357/acf2f4}

\bibitem[{{Wendeborn} {et~al.}({submitted}){Wendeborn}, {Espaillat}, {Lopez},
  {Thanathibodee}, {Robinson}, {Pittman}, \& {Calvet}}]{PaperI}
{Wendeborn}, J., {Espaillat}, C.~C., {Lopez}, S., {et~al.} {submitted}, \apj

\bibitem[{{Wendeborn} {et~al.}({in prep.}){Wendeborn}, {Espaillat},
  {Thanathibodee}, {Robinson}, {Pittman}, \& {Calvet}}]{PaperIII}
{Wendeborn}, J., {Espaillat}, C.~C., {Thanathibodee}, T., {et~al.} {in prep.},
  \apj

\bibitem[{{Zsidi} {et~al.}(2022){Zsidi}, {Manara}, {K{\'o}sp{\'a}l}, {Hussain},
  {{\'A}brah{\'a}m}, {Alecian}, {B{\'o}di}, {P{\'a}l}, \& {Sarkis}}]{Zsidi2022}
{Zsidi}, G., {Manara}, C.~F., {K{\'o}sp{\'a}l}, {\'A}., {et~al.} 2022, \aap,
  660, A108, \dodoi{10.1051/0004-6361/202142203}

\end{thebibliography}

\appendix

\section{Photometry Scaling} \label{appendix: Scaling}

Due to differences in calibration between the various sources of photometry, we scale our photometry to achieve more consistency. In some cases we also scale one bandpass to another similar bandpass, like Johnson $R$ and $I$ to Sloan $r$ and $i$, respectively. Our process for scaling the various sources of photometry for a given bandpass is as follows: 

\begin{enumerate}
    \item We first select a source of photometry to be used as a baseline to which other sources will be scaled. This baseline source was typically our LCOGT photometry, as it is generally the most extensive. In some cases (when insufficient LCOGT photometry was present) we use AAVSO photometry as our baseline.
    \item Next, we select observations from each source that are contemporaneous to within 2 hours. We assume that no significant variability occurs within 2 hours.
    \item Using these contemporaneous data, we fit a simple line to the calibrated magnitudes from each source, with slope $m$ and intercept $b$.
    \item Provided there are at least 3 contemporaneous points and that the $r$ fit coefficient is at least 0.75 (indicating a good linear fit), we use these linear relationships to scale $all$ of the non-baseline photometry, not just the data that are contemporaneous.
    \item In cases where the linear fit is poor or not enough contemporaneous photometry exists, we use multiple sources for scaling.
    \begin{itemize}
        \item For example, no $V$ photometry from Konkoly is contemporaneous with LCOGT. However, it is contemporaneous with AAVSO, and AAVSO is contemporaneous with LCOGT, allowing us to scale Konkoly to LCOGT using two linear relationships.
    \item In a few cases, like for most of the $B$-band photometry, no sources are contemporaneous. In these cases, we simply scale one source to the other using the median flux.
    \end{itemize}
\end{enumerate}

\begin{figure*}
    \centering
    \includegraphics[width=0.85\textwidth]{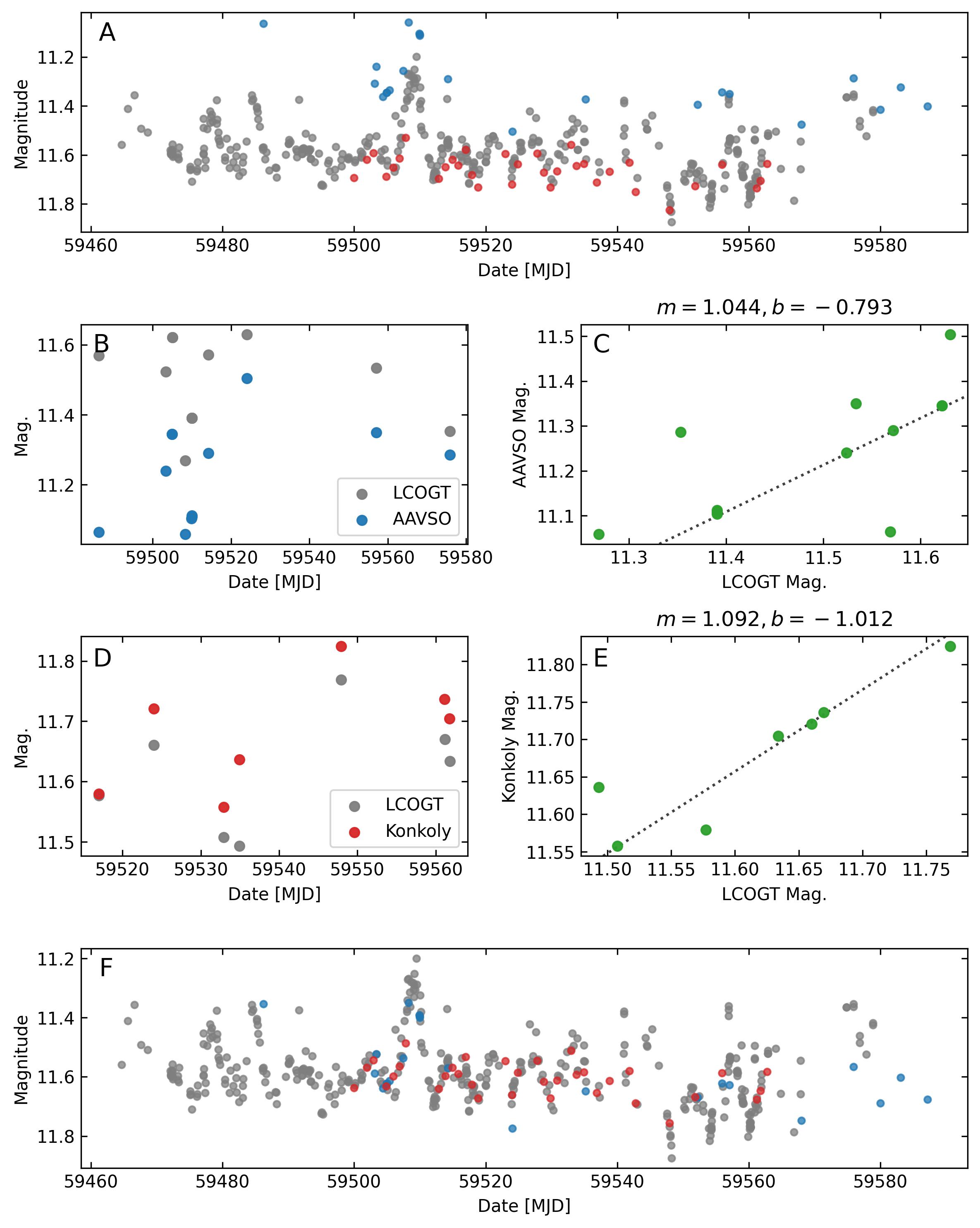}
    \caption{An example visualizing how we scale various sources of photometry. A: Raw, un-scaled $r$-band light curves of GM~Aur E1 from LCOGT (grey), AAVSO (blue), and Konkoly (red). Notice that the data do not necessarily overlap. B: Contemporaneous (within 2 hours) data from LCOGT and AAVSO. C: Linear fit between the contemporaneous data from panel B, with slope $m$ and intercept $b$ noted. D: Contemporaneous (within 2 hours) data from LCOGT and Konkoly. E: Linear fit between the contemporaneous data from panel D, with slope $m$ and intercept $b$ noted. F: Scaled light curves using the linear fits from panels C and E.}
    \label{fig: Scaling}
\end{figure*}

\begin{deluxetable*}{c | c c c | c}[htp]
\tablecaption{Linear Scaling Relationships \label{tab: Scaling}}
\centering
\tablehead{
\colhead{Source} & \colhead{$m$} & \colhead{$b$} & \colhead{$r$} & \colhead{Scaled To}
}
\startdata
\hline
\multicolumn{5}{c}{TW~Hya} \\
\hline
$V_{\mathrm{AAVSO}}$ & 1.16 & -1.57 & 0.90 & $V_{\mathrm{LCOGT}}$ \\
$R_{\mathrm{AAVSO}}$ & 1.06 & -0.80 & 0.96 & $r_{\mathrm{LCOGT}}$ \\
$I_{\mathrm{AAVSO}}$ & 1.07 & -1.06 & 0.90 & $i_{\mathrm{LCOGT}}$ \\
$TESS$ & 1.76 & -7.90 & 0.73 & $i_{\mathrm{LCOGT}}$, $I_{\mathrm{AAVSO}}$ \\
\hline \hline 
\multicolumn{5}{c}{RU~Lup} \\
\hline
$g_{\mathrm{ASAS-SN}}$ & 1.33 & -3.64 & 0.78 & $g_{\mathrm{LCOGT}}$ \\
$V_{\mathrm{AAVSO}}$ & 1.27 & -2.77 & 0.88 & $V_{\mathrm{LCOGT}}$ \\
$R_{\mathrm{AAVSO}}$ & 1.04 & -0.69 & 0.85 & $r_{\mathrm{LCOGT}}$ \\
$I_{\mathrm{AAVSO}}$ & 0.99 & -0.50 & 0.86 & $i_{\mathrm{LCOGT}}$ \\
\hline \hline 
\multicolumn{5}{c}{BP~Tau} \\
\hline
$g_{\mathrm{ASAS-SN}}$ & 0.97 & 0.46 & 0.97 & $g_{\mathrm{LCOGT}}$ \\
$g_{\mathrm{ZTF}}$ & 1.07 & -0.84 & 0.94 & $g_{\mathrm{LCOGT}}$ \\
$V_{\mathrm{AAVSO}}$ & 1.32 & -3.67 & 0.84 & $V_{\mathrm{LCOGT}}$ \\
$R_{\mathrm{AAVSO}}$ & 1.07 & -1.19 & 0.82 & $r_{\mathrm{LCOGT}}$ \\
$r_{\mathrm{Konkoly}}$ & 0.72 & 3.23 & 0.89 & $r_{\mathrm{LCOGT}}$ \\
$R_{\mathrm{CrAO}}$ & 0.97 & -0.03 & 0.98 & $r_{\mathrm{LCOGT}}$ \\
$I_{\mathrm{AAVSO}}$ & 1.43 & -5.44 & 0.76 & $i_{\mathrm{LCOGT}}$ \\
$i_{\mathrm{Konkoly}}$ & 1.14 & -1.64 & 0.89 & $i_{\mathrm{LCOGT}}$ \\
$I_{\mathrm{CrAO}}$ & 0.85 & 1.00 & 0.99 & $i_{\mathrm{LCOGT}}$ \\
$TESS$ & 3.08 & -22.94 & 0.60 & $i_{\mathrm{LCOGT}}$ \\
\hline \hline 
\multicolumn{5}{c}{GM~Aur} \\
\hline
$g_{\mathrm{ASAS-SN}}$ & 0.98 & -1.29 & 0.95 & $g_{\mathrm{LCOGT}}$ \\
$g_{\mathrm{ZTF}}$ & 1.12 & -1.47 & 0.86 & $g_{\mathrm{LCOGT}}$ \\
$V_{\mathrm{AAVSO}}$ & 0.46 & 6.47 & 0.96 & $V_{\mathrm{LCOGT}}$ \\
$V_{\mathrm{Konkoly}}$ & 0.65 & 4.11 & 0.76 & $V_{\mathrm{LCOGT}}$, $V_{\mathrm{AAVSO}}$ \\
$R_{\mathrm{AAVSO}}$ & 1.04 & -0.79 & 0.98 & $r_{\mathrm{LCOGT}}$ \\
$r_{\mathrm{Konkoly}}$ & 1.09 & -1.01 & 0.85 & $r_{\mathrm{LCOGT}}$ \\
$R_{\mathrm{CrAO}}$ & 1.18 & -2.32 & 0.78 & $r_{\mathrm{LCOGT}}$ \\
$I_{\mathrm{AAVSO}}$ & 1.26 & -3.42 & 0.88 & $i_{\mathrm{LCOGT}}$ \\
$i_{\mathrm{Konkoly}}$ & 1.07 & -0.72 & 0.89 & $i_{\mathrm{LCOGT}}$ \\
$I_{\mathrm{CrAO}}$ & 0.63 & 3.59 & 0.79 & $i_{\mathrm{LCOGT}}$ \\
$TESS$ & 1.26 & -2.73 & 0.89 & $i_{\mathrm{LCOGT}}$ \\
\enddata
\end{deluxetable*}
\clearpage

\onecolumngrid
\section{Further Analysis of Light Curve Periodicity} \label{appendix: Periodicity}

The periodograms presented in Section~\ref{sec: Analysis and Results} are created using the Lomb-Scargle periodogram, a well-known period-finding algorithm in the field of astronomy. Its most useful characteristic is the ability to analyze light curves with gaps and irregular spacing, necessary for ground-based observations. It does, however, come with some considerations that one must account for to properly interpret its output \citep{VanderPlas2018}. In this section we consider some of these specifics and how they may impact the periods determined in Section~\ref{sec: Analysis and Results}. 

\subsection{Aliased Frequencies} \label{appendix: Aliases}

\citet{VanderPlas2018} note that several spurious, non-physical frequencies can be found in Lomb-Scargle periodograms, notably:

\begin{itemize}
    \item Peaks near $f=$\fpeak/m, where m$\in \{2,3\}$ and \fpeak is a strong, observed peak in the periodogram. These are known as $m$-harmonics.
    \item Peaks near $f=|$\fpeak$\pm n$\df$\|$, where n$\in \{1,2\}$ and \df is a characteristic frequency of the window function (see below). We refer to these as aliases.
\end{itemize}

$m$-harmonics are sometimes present our light curves, typically when there is a clear, well-defined periodic signal. This includes GM~Aur E1, where the 6-day period is strong. A corresponding period near 12 days is seen in each bandpass. These harmonics may also be seen in TW~Hya E2, where signal is seen at $\sim$3.5 and $\sim$7.1 days in most bandpasses. These $m$-harmonics may also be present in the low-power/noisy signals seen in several epochs (TW~Hya E1, BP~Tau E1/E2) below about 8 days, though determining which is the true signal is difficult.

\begin{figure*}
    \centering
    \includegraphics[width=0.85\textwidth]{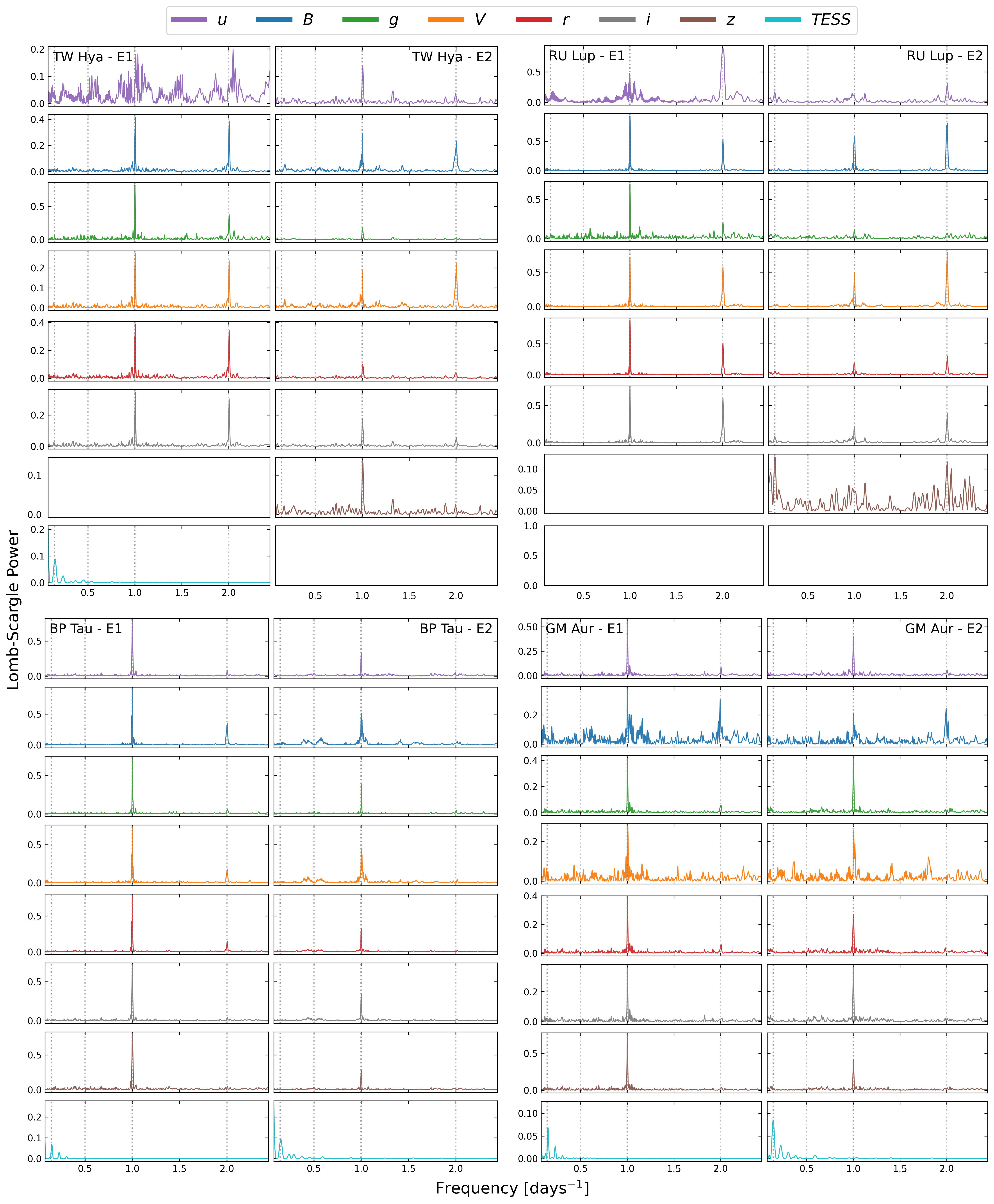}
    \caption{Window function power spectra for our light curves. Gray dashed lines denote frequencies of 1/7, 0.5, 1, and 2 day$^{-1}$ that are commonly found.}
    \label{fig: Window Power}
\end{figure*}

Figure \ref{fig: Window Power} shows the window function power spectrum of our light curves. These are created by essentially taking the Lomb-Scargle periodogram where all the data is set to 1. Many of our light curves possess window power (sharp peaks) at 1 day$^{-1}$ and related harmonics like 2 day$^{-1}$ and 0.5 day$^{-1}$. Our $TESS$ light curves also show power near 1/7 day$^{-1}$. Given that \df=1 day$^{-1}$, signals that fall at \fpeak$\pm$n\df should be ignored.

\begin{figure*}
    \centering
    \includegraphics[width=0.95\textwidth]{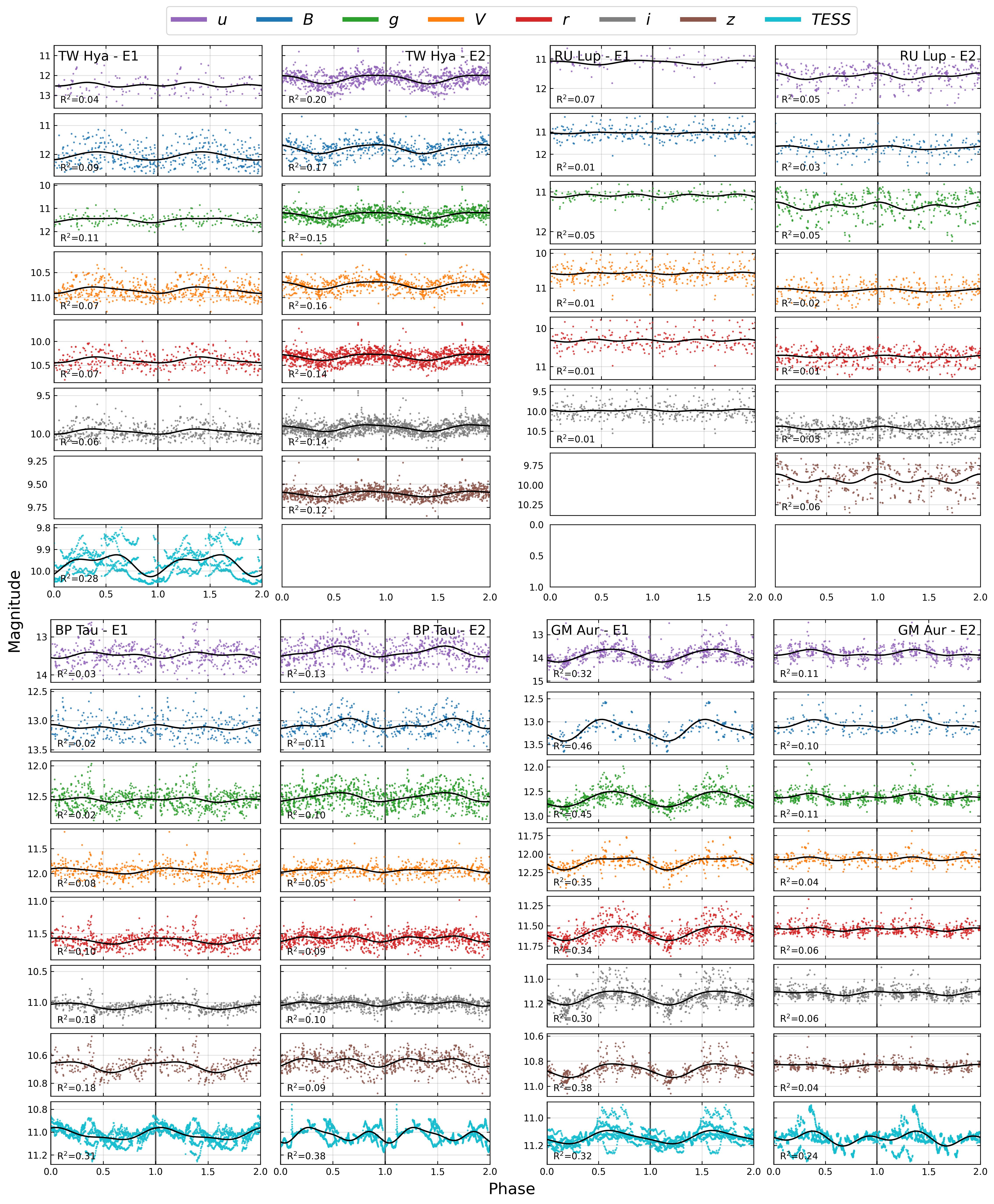}
    \caption{Phase-folded light curves for each target and filter using the periods identified in Section~\ref{sec: Analysis and Results}: 3.57/3.54 days, 3.71/3.71 days, 8.22/8.31 days, and 6.01/6.00 days for E1/E2 in TW~Hya, RU~Lup, BP~Tau, and GM~Aur, respectively. The solid black line is the GP fitted curve. The Coefficient of Determination (R$^2$) is also calculated between each folded light curve its respective GP fit, and is shown in the lower-left corner of each panel.}
    \label{fig: Phase Folded}
\end{figure*}

We attempt to account for these aliased frequencies more robustly by first removing the primary periods detected in our periodograms. These are 3.57/3.54 days, 3.71/3.71 days, 8.22/8.31 days, and 6.01/6.00 days for E1/E2 in TW~Hya, RU~Lup, BP~Tau, and GM~Aur identified in Section~\ref{sec: Analysis and Results}. We do this in a similar manner to that of determining the $Q$ periodicity parameter (see Section~\ref{sec: QM Results}). In essence, we fit a Gaussian Process (GP) to a phase-folded light curve and subtract this signal from the original light curve. Figure~\ref{fig: Phase Folded} shows the light curves phase-folded according to the above periods along with the fitted GP curve. In cases where the primary period in clearly detected (TW~Hya, E2, all filters; GM~Aur, E1, all filters; BP~Tau, E1, $riz$; BP~Tau, E2, $uBg$), there is sinusoidal structure across one phase in the GP fit (and higher R$^2$ values), suggesting this period is representative of the rotation during that epoch. Additionally, the structure in various filters is typically similar and in phase. In other cases, the folded light curves show no structure (with correspondingly low R$^2$), reinforcing that little to no periodicity is seen at the chosen period.

\begin{figure*}
    \centering
    \includegraphics[width=0.85\textwidth]{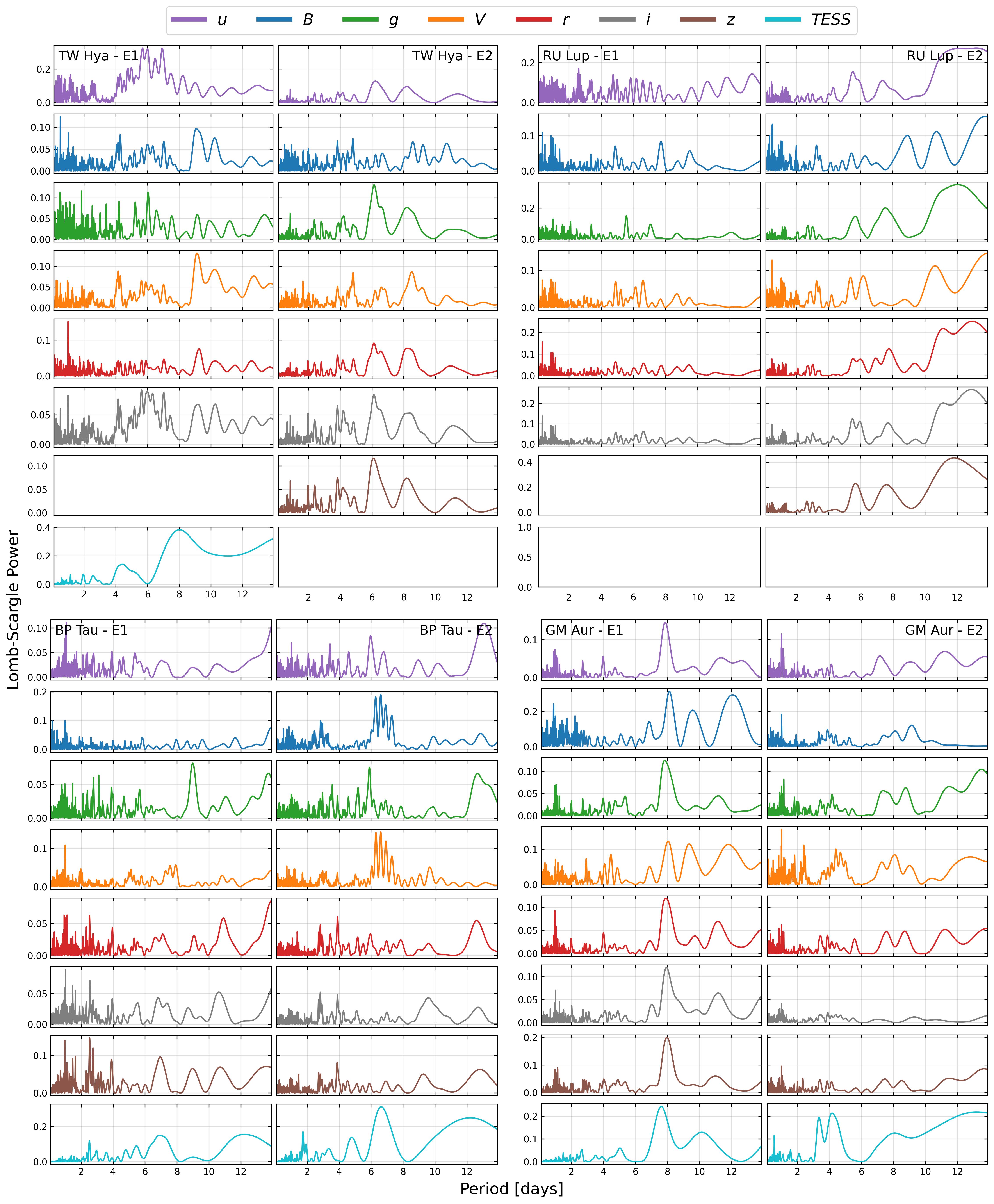}
    \caption{Lomb-Scargle periodograms for all light curves after removing the primary periodicity via Gaussian Process. }
    \label{fig: Periods 2}
\end{figure*}

Next, we perform the Lomb-Scargle periodogram as in Section~\ref{sec: Analysis and Results}, this time on the primary-period subtracted light curves. These periodograms are shown in Figure \ref{fig: Periods 2}. Peaks in these periodograms should be real and not related to aliasing of the primary period.

\end{document}